\documentclass[traditabstract]{aa}
\usepackage{amsmath}
\usepackage{amssymb}
\usepackage{booktabs}
\usepackage{multirow}
\usepackage{nicefrac}
\usepackage{txfonts}
\usepackage{epsfig,graphicx}
\usepackage{units}
\usepackage[colorlinks=true, linkcolor=blue, citecolor=blue, urlcolor=blue]{hyperref}
\usepackage{natbib}
\bibpunct{(}{)}{;}{a}{}{,}  
\newcommand{\txd}{\text{d}}

\newcommand{\bfk}{\boldsymbol{k}}
\newcommand{\bfknew}{\bfk_\text{new}}
\newcommand{\Inew}{I_\text{new}}
\newcommand{\bfn}{\boldsymbol{n}}
\newcommand{\bfnnew}{\bfn_\text{new}}
\newcommand{\bfv}{\boldsymbol{v}}
\newcommand{\bfvnew}{\bfv_\text{new}}

\newcommand{\bfd}{\boldsymbol{d}}
\newcommand{\bfS}{\boldsymbol{S}}
\newcommand{\bfSnew}{\bfS_\text{new}}

\newcommand{\mum}[1]{\unit[#1]{\ensuremath{\mu}m}}
\newcommand{\kpc}[1]{\unit[#1]{kpc}}
\newcommand{\AU}[1]{\unit[#1]{AU}}
\newcommand{\degree}{\ensuremath{^\circ}}

\begin{document}

\title{Polarization in Monte Carlo radiative transfer and dust\\scattering polarization signatures of spiral galaxies}

\titlerunning{Polarization in RT simulations}

\author{C. Peest\thanks{christian.peest@ugent.be}\inst{1,2}
    \and P. Camps\inst{1}
    \and M. Stalevski\inst{1,3,4}
    \and M. Baes\inst{1}
    \and R. Siebenmorgen\inst{2}}

\institute{Sterrenkundig Observatorium, Universiteit Gent, Krijgslaan 281 S9, B-9000 Gent, Belgium
    \and European Southern Observatory, Karl-Schwarzschild-Str. 2, D-85748 Garching b. M\"unchen, Germany
    \and Universidad de Chile, Observatorio Astronomico Nacional Cerro Calan, Camino El Observatorio 1515, Santiago, Chile
    \and Astronomical Observatory, Volgina 7, 11060 Belgrade, Serbia}

\abstract{Polarization is an important tool to further the understanding of interstellar dust and the sources behind it. In this paper we describe our implementation of polarization that is due to scattering of light by spherical grains and electrons in the dust Monte Carlo radiative transfer code SKIRT. In contrast to the implementations of other Monte Carlo radiative transfer codes, ours uses co-moving reference frames that rely solely on the scattering processes. It fully supports the peel-off mechanism that is crucial for the efficient calculation of images in 3D Monte Carlo codes. We develop reproducible test cases that push the limits of our code. The results of our program are validated by comparison with analytically calculated solutions. Additionally, we compare results of our code to previously published results. We apply our method to models of dusty spiral galaxies at near-infrared and optical wavelengths. We calculate polarization degree maps and show them to contain signatures that trace characteristics of the dust arms independent of the inclination or rotation of the galaxy.}


\maketitle


\section{Introduction}
Many astronomical objects contain or are shrouded by dust.
Often, a non-negligible fraction of ultraviolet (UV) to near-infrared (NIR) radiation emitted by embedded sources is scattered or absorbed by dust grains before leaving the system.
Scattered radiation is generally polarized.
The polarization state of the light can be used to deduce information about the grains that would not be available using intensity measurements alone \citep{Scicluna2015}.
There are indications that dust properties differ widely and systematically between galaxies \citep{Fitzpatrick1990,Gordon2003,Remy2015,Dale2012,DeVis2016} and that they can vary significantly within a galaxy \citep{Draine2014,Mattsson2014}. Polarimetric studies can help in constraining these properties.
Theoretical frameworks for modeling radiative transfer therefore usually include a section on polarization \citep[see, e.g.,][]{Chandrasekhar1950, VanDeHulst1957}.

Numerical simulations of dust radiative transfer most commonly use the Monte Carlo technique \citep[see, e.g.,][]{Whitney2011,Steinacker2013}.
Codes using this method track many individual photon packages as they propagate through the dusty medium, simulating emission, scattering, and absorption events based on random variables drawn from the appropriate probability distributions.
While it is conceptually straightforward to track the polarization state of a photon package as part of this process, there are many details to be considered, and the implementation complexity depends on the assumptions and approximations one is willing to make.
Moreover, the dust model used by the code must provide the extra properties necessary to calculate the changes to the polarization state for each interaction with a dust grain.

As a result, various authors have made different choices for implementing polarization in Monte Carlo radiative transfer (MCRT) codes.
Most commonly, the MCRT codes consider only scattering by spherical dust grains \citep[e.g.,][]{Bianchi1996,Pinte2006,Min2009,Robitaille2011,Goosmann2014}.
Some codes include more advanced support for polarization by absorption and scattering off aligned spheroids \citep{Whitney2002,Lucas2003,Reissl2016} and/or for polarized dust emission \citep{Reissl2016}.

To verify the correctness of the various polarization implementations, authors sometimes compare the results between codes \citep[e.g.][]{Pinte2009}.
Because of the variations in assumptions and capabilities, however, such a comparison is tricky and the `correct' result is usually simply assumed to be the result obtained by a majority of the codes.
Even when the basic assumptions about grain shape and alignment as well as the dust mixture are the same and the codes support the same polarization processes, comparing results is usually complicated.

In this paper we present a robust framework that is independent
of a coordinate system for implementing polarization in a tree-dimensional (3D) MCRT code. The mathematical formulation and the numerical calculations in our method rely solely on reference frames determined by the physical processes under study (i.e., the propagation direction or the scattering plane) and not on those determined by the coordinate system (i.e., the $z$-axis).
This approach avoids numerical instabilities for special cases (i.e., a photon package propagating in the direction of the $z$-axis or close to it) and enables a more streamlined implementation.

We have implemented this framework in SKIRT\footnote{http://www.skirt.ugent.be} \citep{Baes2003, Baes2011, Camps2015a}, a versatile multipurpose Monte Carlo dust radiative transfer code. It has been designed and optimized for systems with a complex 3D structure, as multiple components are configured separately and construct a more complex model for the dust and/or radiation sources \citep{Baes2015}. The code is equipped with a range of efficient grid structures on which the dust can be spatially discretized, including octree, k-d tree, and Voronoi grids \citep{Saftly2013,Saftly2014,Camps2013}. A powerful hybrid parallelization scheme has been developed that guarantees an optimal speed-up and load balancing \citep{Verstocken2016}, and it opens up a wide range of possible polarization applications.
In order to test the correct behavior and the accuracy of our implementation, we have developed a number of analytical test cases designed to validate polarization implementations in a structured manner.
Furthermore, we carefully match our polarimetric conventions to the recommendations issued by the International Astronomical Union \citep{IAU1974}.

In large-scale dust systems complex geometries arise and need to be handled by the codes. We first apply our method to some elementary models of dusty disk galaxies, enabling a qualitative comparison with the two-dimensional (2D) models of \citet{Bianchi1996}.
We also perform the polarization part of the \citet{Pinte2009} benchmark and compare with the published results.

We then implement spiral arms into dusty disk galaxy models and show that this produces a marked polarimetric signature tracing the positions of the arms.
Our current implementation supports only scattering by spherical grains. Dichroic extinction and more complex grain shapes may also have a strong influence \citep{Voshchinnikov2012,Siebenmorgen2014,Draine2009} and will be supported in future work.

In Sect.~\ref{sec:Polarization} we summarize the notation and conventions used in this paper to describe the polarization state of electromagnetic radiation, and we provide recipes for translation into other conventions.
We then present our method and its implementation in Sect.~\ref{sec:Method}, and the accompanying analytical test cases and their results in Sect.~\ref{sec:Validation}. The application of our method to benchmark tests is described in Sect.~\ref{sec:Models}. The dusty spiral galaxy model is described and implemented in Sect.~\ref{sec:spiralarmmodel}. We summarize and conclude in Sect.~\ref{sec:Conclusion}.


\section{Polarization}\label{sec:Polarization}

\begin{figure}
\centering
\includegraphics[width=0.7\columnwidth]{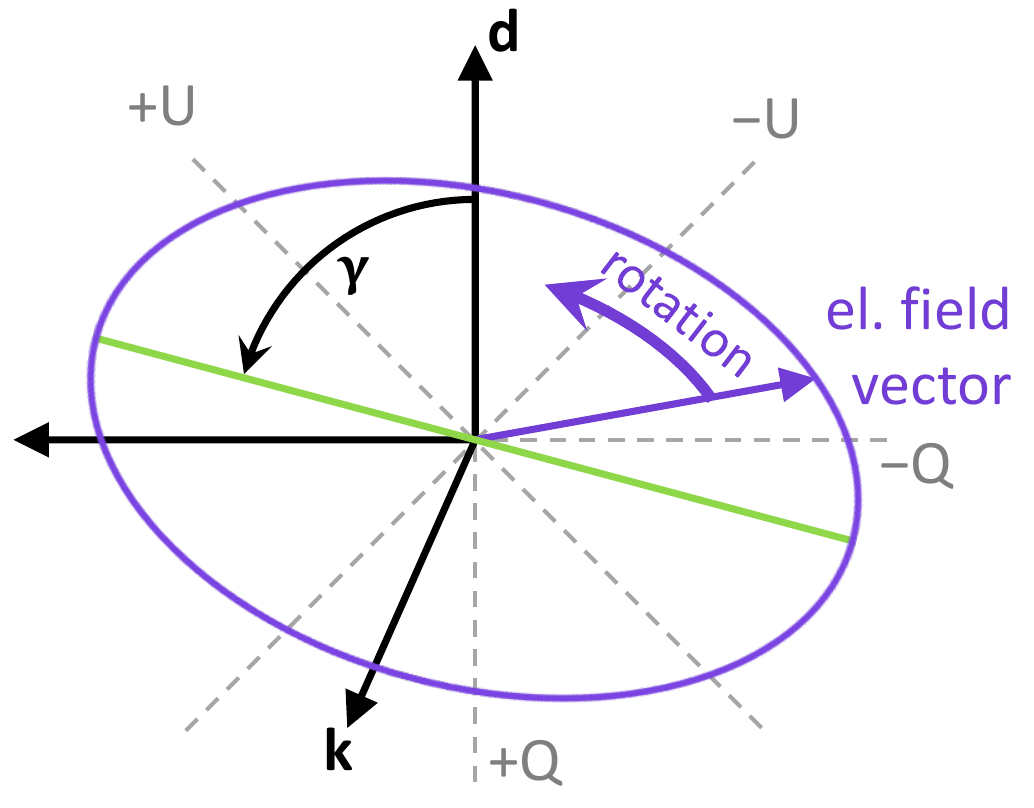}
\caption{Illustration of the Stokes vector conventions recommended by \citet{IAU1974} and used in this paper.
The radiation beam travels along its propagation direction, $\bfk$, out of the page.
The electric field vector describes an ellipse over time.
The linear polarization angle, $\gamma$, is given by the angle between the primary axis of the ellipse (green line) and the reference direction, $\bfd$.
The position angle of the electric field vector increases with time, the beam has right-handed circular polarization.}
\label{fig:StokesEllipse}
\end{figure}

\subsection{Stokes vector}\label{sec:StokesVector}
The polarization state of electromagnetic radiation is commonly described by the Stokes vector, $\bfS$ \citep[see, e.g.,][]{VanDeHulst1957,Chandrasekhar1950,Bohren1998,Mishchenko1999},
\begin{equation}
\bfS = \begin{pmatrix} I \\ Q \\ U \\ V \end{pmatrix},
\end{equation}
where $I$ represents the intensity of the radiation, $Q$ and $U$ describe linear polarization, and $V$ describes circular polarization.
The degrees of total and linear polarization, $P$ and $P_{\text{L}}$, can be written as a function of the Stokes parameters,
\begin{align}
P &= \sqrt{Q^2 + U^2+V^2}/I,\\
P_{\text{L}} &= \sqrt{Q^2+U^2}/I\label{eq:degreeOfLinearPolarisation}.
\end{align}
The (linear) polarization angle, $\gamma$, can be written as
\begin{equation}\label{eq:directionOfPolarisation}
\gamma = \frac12\arctan_2\left(\frac{U}{Q}\right),
\end{equation}
where $\arctan_2$ denotes the inverse tangent function that preserves the quadrant.
Combining Eqs.~\eqref{eq:degreeOfLinearPolarisation} and~\eqref{eq:directionOfPolarisation}, we can also write
\begin{subequations}\label{polar}
\begin{align}
Q &= I P_{\text{L}} \cos2\gamma,\\
U &= I P_{\text{L}} \sin2\gamma.
\end{align}
\end{subequations}

The values of $Q$ and $U$ depend on the polarization angle $\gamma$, which describes the angle between the direction of linear polarization and a given reference direction, $\bfd$, in the plane orthogonal to the propagation direction, $\bfk$.
The angle is measured counter-clockwise when looking at the source, as illustrated in Fig.~\ref{fig:StokesEllipse}.
A linear polarization angle in the range $0<\gamma<\pi/2$ implies a positive $U$ value.
A radiation beam is said to have right-handed circular polarization (with $V>0$) when the electric field vector position angle increases with time, and left-handed when it decreases.

The reference direction, $\bfd$, can be chosen arbitrarily as long as it is well defined and perpendicular to the propagation direction.
However, when the polarization state changes as a result of an interaction (e.g., a scattering event), most recipes for properly adjusting the Stokes vector require the reference direction to have a specific orientation (e.g., lying in the scattering plane).
Before applying the recipe, the existing reference direction must be rotated about the propagation direction to match this requirement.
This is accomplished by multiplying the Stokes vector by a rotation matrix, ${\bf{R}}(\varphi)$,
\begin{equation}
\bfSnew = {\bf{R}}(\varphi)\,\bfS.
\label{eq:applyRotation}
\end{equation}
A rotation about the direction of propagation by an angle $\varphi$, counter-clockwise when looking toward the source of the beam, is described by the matrix
\begin{equation}
{\bf{R}}(\varphi)
=
\begin{pmatrix}
1 & 0 & 0 & 0 \\
0 & \cos2\varphi & \sin2\varphi & 0 \\
0 & -\sin2\varphi & \cos2\varphi & 0 \\
0 & 0 & 0 & 1
\end{pmatrix}.
\label{eq:rotation}
\end{equation}

To record the polarization state change for a scattering event, the Stokes vector is multiplied by the M\"uller matrix, ${\bf{M}}$, corresponding to the event, assuming that the reference direction lies in the scattering plane (as well as in the plane orthogonal to the propagation direction).
The M\"uller matrix components depend on the geometry of the scattering event and the physical properties of the scatterer, and they often depend on the wavelength.
In general, the M\"uller matrix is
\begin{equation}\label{eq:generalMueller}
{\bf{M}(\theta,\lambda)}
=
\begin{pmatrix}
S_{11} & S_{12} & S_{13} & S_{14} \\
S_{21} & S_{22} & S_{23} & S_{24} \\
S_{31} & S_{32} & S_{33} & S_{34} \\
S_{41} & S_{42} & S_{43} & S_{44}
\end{pmatrix},
\end{equation}
where $\theta$ is the angle between the propagation directions before and after the scattering event, and $\lambda$ is the wavelength of the radiation.
For clarity of presentation, we drop the dependencies from the notation for the individual M\"uller matrix components.
Including the reference direction adjustments before and after the actual scattering event, $\varphi$ and $\varphi_\text{new}$, the transformation of a Stokes vector for a scattering event can thus be written as\begin{equation}\label{eq:applyMueller}
\bfSnew = \bf{R}(\varphi_\text{new})\,\bf{M}(\theta,\lambda)\,\bf{R}(\varphi)\,\bfS.
\end{equation}

For scattering by spherical particles, the M\"uller matrix simplifies to \citep{Kruegel2002}
\begin{equation}\label{eq:sphereScattering}
{\bf{M_\mathrm{Sph}}}(\theta,\lambda)
=
\begin{pmatrix}
S_{11} & S_{12} & 0 & 0 \\
S_{12} & S_{11} & 0 & 0 \\
0 & 0 & S_{33} & S_{34} \\
0 & 0 & -S_{34} & S_{33}
\end{pmatrix},
\end{equation}
again assuming that the reference direction lies in the scattering plane.
The M\"uller matrices for a particular grain size and material can be calculated using Mie theory \citep[see e.g.,][]{Voshchinnikov1993, Bohren1998, Pena2009}.

For scattering by electrons, also called Thomson scattering, the M\"uller matrix is wavelength-independent and can be expressed analytically as a function of the scattering angle  \citep{Bohren1998},
\begin{equation}\label{eq:ThomsonScattering}
{\bf{M_\mathrm{Th}}}(\theta)
=\frac12
\begin{pmatrix}
\cos^2\theta +1 & \cos^2\theta -1 & 0 & 0 \\
\cos^2\theta -1 & \cos^2\theta +1 & 0 & 0 \\
0 & 0 & 2\cos\theta & 0 \\
0 & 0 & 0 &  2\cos\theta
\end{pmatrix}.
\end{equation}

\subsection{Conventions}\label{sec:StokesConventions}

\begin{table}
\centering
\caption{Conventions adopted by various authors regarding the sign of the Stokes parameters $U$ and $V$ relative to \citet{IAU1974} ($+U$,$+V$).}
\label{tab:stokesConventions}
\begin{tabular}{r|l|l}
        \toprule\toprule
\multicolumn{1}{c|}{ } & \multicolumn{1}{c|}{$+U$}  & \multicolumn{1}{c}{$-U$}  \\
\hline
 \multirow{8}{*}{$+V$}  & \citet{IAU1974}   & \citet{Chandrasekhar1950} \\
                    & \citet{Martin1974}    & \citet{VanDeHulst1957}    \\
                    & \citet{Tsang1985}     & \citet{Hovenier1983}*     \\
                    & \citet{Trippe2014}    & \citet{Fischer1994}*      \\
                    &                       & \citet{Code1995}*         \\
                    &                       & \citet{Mishchenko1999}*   \\
                    &                       & \citet{Gordon2001}*       \\
                    &                       & \citet{Lucas2003}         \\
                    &                       & \citet{Gorski2005}        \\
\hline
\multirow{2}{*}{$-V$}       & \citet{Shurcliff1962} & \citet{Bohren1998}\\
                    & \citet{Bianchi1996}*  & \citet{Mishchenko2002}    \\
        \bottomrule
\multicolumn{3}{c}{* Convention indicated by citing papers with this convention}
\end{tabular}
\end{table}

In this paper we define the Stokes vector following the recommendations of the International Astronomical Union \citep{IAU1974}, as presented in Sect.~\ref{sec:StokesVector} and illustrated in Fig.~\ref{fig:StokesEllipse}.
Historically, however, authors have used various conventions for the signs of the Stokes parameters $U$ and $V$ \citep[][see also a recent IAU announcement\footnote{http://iau.org/static/archives/announcements/pdf/ann16004a.pdf}]{Hamaker1996}.
For example, the polarization angle $\gamma$ is sometimes measured while looking toward the observer rather than toward the source, flipping the sign of both $U$ and $V$.
Reversing the definition of circular polarization handedness also flips the sign of $V$.
Table~\ref{tab:stokesConventions} lists a number of references with the conventions adopted by the authors.

Assuming that the adopted conventions are properly documented, translating the values of the Stokes parameters from one convention into another is straightforward -- by flipping the signs appropriately.
When comparing or mixing formulas and recipes constructed for different conventions,
changes in the signs of $U$ and $V$ affect the sign of the M\"uller matrix components (Eq.~\ref{eq:generalMueller}) in the third row and column and fourth row and column, respectively.
Mathematically this can be described by multiplying the M\"uller matrix by signature matrices,
\begin{align}
{\bf{M_\mathrm{+U,+V}}}
&=
\begin{pmatrix}
1 & 0 & 0 & 0 \\
0 & 1 & 0 & 0 \\
0 & 0 & \sigma & 0 \\
0 & 0 & 0 & \varsigma
\end{pmatrix}
{\bf{M_\mathrm{\sigma U, \varsigma V}}}
\begin{pmatrix}
1 & 0 & 0 & 0 \\
0 & 1 & 0 & 0 \\
0 & 0 & \sigma & 0 \\
0 & 0 & 0 & \varsigma
\end{pmatrix}\\
&=
\begin{pmatrix}
S_{11} & S_{12} & \sigma S_{13} & \varsigma S_{14} \\
S_{21} & S_{22} & \sigma S_{23} & \varsigma S_{24} \\
\sigma S_{31} & \sigma S_{32} & S_{33} & \sigma \varsigma S_{34} \\
\varsigma S_{41} & \varsigma S_{42} & \sigma \varsigma S_{43} & S_{44}
\end{pmatrix} ,
\end{align}
with $\sigma$ and $\varsigma$ being $+1$ or $-1$. In case of the rotation matrix, Eq.~\eqref{eq:rotation}, this implies that the signs in front of the sine functions change based on the chosen convention.


\section{Method}\label{sec:Method}

\subsection{SKIRT code}\label{sec:SKIRT}

SKIRT \citep{Baes2003, Baes2011, Camps2015a} is a public multipurpose MCRT code for simulating the effect of dust on radiation in astrophysical systems.
It offers full treatment of absorption and multiple anisotropic scattering by the dust, self-consistently computes the temperature distribution of the dust and the thermal dust reemission, and supports stochastic heating of dust grains \citep{Camps2015b}.
The code handles multiple dust mixtures and arbitrary 3D geometries for radiation sources and dust populations, including grid- or particle-based representations generated by hydrodynamical simulations \citep{Camps2016}.

SKIRT is predominantly used to study dusty galaxies \citep{Baes2010, DeLooze2012, DeLooze2014, DeGeyter2014, DeGeyter2015, Saftly2015, Mosenkov2016, Viaene2016b}, but it has also been applied to active galactic nuclei \citep{Stalevski2012a,Stalevski2016}, molecular clouds \citep{Hendrix2015}, and binary systems \citep{Deschamps2015, Hendrix2016}.

Employing the MCRT technique, SKIRT represents electromagnetic radiation as a sequence of discrete photon packages.
Each photon package carries a specific amount of energy (luminosity) at a given wavelength.
A SKIRT simulation follows the individual paths of many photon packages as they propagate through the dusty medium.
The photon package life cycle is governed by various events determined stochastically by drawing random numbers from the appropriate probability distributions.
A photon package is created at a random position based on the luminosities of the sources and is emitted in a random direction depending on the (an)isotropy of the selected source.
Depending on the dust material properties and spatial distribution, the photon package then undergoes a number of scattering events at random locations \citep[using forced scattering; see][]{Cashwell1959}, and is attenuated by absorption along its path \citep[using continuous absorption; see][]{Lucy1999, Niccolini2003}.

To boost the efficiency of the simulation and reduce the noise levels at the simulated observers, SKIRT employs the common peel-off optimization technique \citep{Yusef1984}.
Rather than waiting until a photon package happens to leave the system under study in the direction of one of the observers, a special photon package is peeled off in the direction of each observer at each emission and scattering event, including an appropriate luminosity bias for the probability that a photon package would indeed be emitted or scattered in that direction.
Meanwhile, the original or main photon package continues its trajectory through the dust until its luminosity has become negligible and the package is discarded.

For the purposes of this paper, we assume that a newly emitted photon package represents unpolarized radiation, and its polarization state is not affected by the attenuation along its path through the dusty medium. We assume that the photon package is scattered by spherical dust grains (which does affect the polarization state).
This leaves us with three types of events: scattering the main photon package into a new direction, peeling off a special photon package toward a given observer, and detecting a peel-off photon package at an observer.
As a first step toward describing the procedures for each of these events, we discuss our approach for handling the Stokes vector reference direction.

\subsection{Reference direction}\label{sec:referencedirection}

\begin{figure}
\centering
\includegraphics[width=0.7\columnwidth, trim={1cm 1.7cm 3cm 2.5cm},clip]{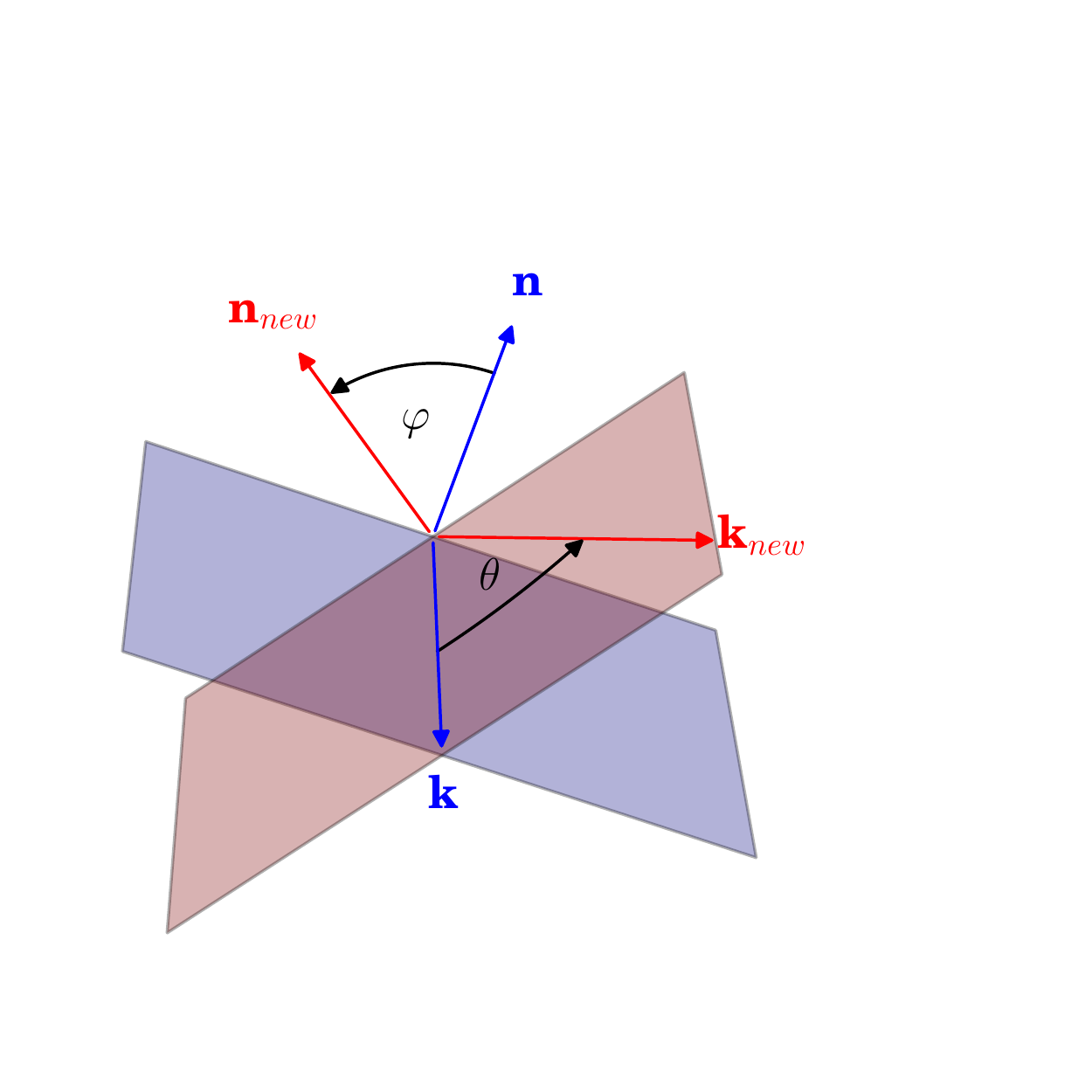}
\caption{Geometry of a scattering event. The angle $\theta$ is between the incoming and outgoing propagation directions $\bfk$ and $\bfknew$. The angle $\varphi$ is given by the normals of the previous and the present scattering planes $\bfn$ and $\bfnnew$.}
\label{PlaneRotation.fig}
\end{figure}

As noted in Sect.~\ref{sec:Polarization}, the Stokes vector describing the polarization state of a photon package is defined relative to a given reference direction, $\bfd$, in the plane perpendicular to the propagation direction.
We define a new direction, $\bfn$, perpendicular to both the propagation direction, $\bfk$, and the reference direction, $\bfd$, which are perpendicular to each other as well, so that\begin{equation}
\bfn = \bfk \times \bfd  \qquad \mathrm{and} \qquad \bfd = \bfn \times \bfk,
\end{equation}
assuming all three vectors are unit vectors.
By definition, the scattering plane contains both the incoming and outgoing propagation directions $\bfk$ and $\bfknew$.
Consider the situation before the event (also, see Fig.~\ref{PlaneRotation.fig}).
If the reference direction $\bfd$, which is always perpendicular to $\bfk$, lies in the scattering plane as well, then $\bfn$ corresponds to the normal of the scattering plane.
A similar situation applies after the scattering event.
We store $\bfn$ rather than $\bfd$ with each photon package, and our procedures below are described in terms of $\bfn$.

Some authors \citep[e.g.,][]{Chandrasekhar1950, Code1995, Gordon2001} choose to rotate the Stokes reference direction in between scattering events into the meridional plane of the coordinate system.
Their procedure uses two rotation operations for each scattering event, one before and one after the event, and requires special care to avoid numerical instabilities when the propagation direction is close to the $z$-axis. The latter occurs because the meridional plane is then ill defined.

We leave the reference direction unchanged after a scattering event and instead perform a single rotation as part of the next scattering event. This method is also applied by \citet{Fischer1994} and \citet{Goosmann2007} and illustrated in Fig.~\ref{PlaneRotation.fig}.
The current scattering plane (red) includes the incoming and outgoing propagation directions $\bfk$ and $\bfknew$, defining the scattering angle $\theta$.
After the previous scattering event, the reference direction has been left in the previous scattering plane (blue), so the angle between the normals $\bfn$ and $\bfnnew$ to the previous and the current scattering planes determines the angle $\varphi$ over which the Stokes vector must be rotated to end up in the current scattering plane.
The transformation of the Stokes vector given in Eq.~\eqref{eq:applyMueller} can therefore be simplified to
\begin{equation}\label{eq:genScatter}
\bfSnew=\bf{M}(\theta,\lambda)\ \bf{R}(\varphi)\ \bfS.
\end{equation}

Care must be taken to properly set a reference normal $\bfn$ for newly emitted photon packages that have not yet experienced a scattering event. Because we assume our sources to emit unpolarized radiation, we can pick any direction perpendicular to the propagation direction. We postpone the details and justification for this procedure to Sect.~\ref{sec:generateReferenceDirection}.

\subsection{Scattering phase function}

The probability that a photon package leaves a scattering event along a particular direction $\bfknew$ for an incoming direction $\bfk$ is given by the phase function, $\Phi(\bfk,\bfknew)\equiv\Phi(\theta,\varphi)$, where $\theta$ and $\varphi$ represent the inclination and azimuthal angles of $\bfknew$ relative to $\bfk$, and where we omit the wavelength dependency from the notation.
In the formulation of Sect.~\ref{sec:Polarization}, we can say that the phase function is proportional to the ratio of the beam intensities, $I$ and $\Inew$, before and after the scattering event,
\begin{equation}
\Phi(\theta,\varphi)
\propto
\frac{\Inew(\theta,\varphi)}{I}.
\end{equation}
For spherical grains, combining Eqs.~\eqref{eq:rotation}, \eqref{eq:sphereScattering}, and~\eqref{eq:genScatter} leads to
\begin{equation}
\Inew(\theta,\varphi) = I S_{11} + S_{12} \left( Q\cos2\varphi+U\sin2\varphi \right)
\end{equation}
and therefore,
\begin{equation}
\Phi(\theta,\varphi)
\propto
S_{11} + S_{12}\left(
\frac{Q}{I}\cos2\varphi+\frac{U}{I}\sin2\varphi
\right).
\end{equation}
Using Equation~(\ref{polar}) and introducing a proportionality factor, $N$, we can write
\begin{equation}\label{phasefunction}
\Phi(\theta,\varphi)
=
N\,S_{11} \left( 1 + P_{\text{L}}\,\frac{S_{12}}{S_{11}}\cos2(\varphi - \gamma) \right).
\end{equation}
The proportionality factor is determined by normalizing the phase function (a probability distribution) to unity. Integration over the unit sphere yields
\begin{align}
N&=\frac{1}{ \int_0^{2\pi} \int_0^{\pi} \left(S_{11} + P_{\text{L}}S_{12}\cos2(\varphi - \gamma)\right)\sin\theta\, \txd\theta\, \txd\varphi }\\
&=\frac{1}{2\pi\int_0^\pi S_{11}\sin\theta\, \txd\theta} \ .
\end{align}

\subsection{Sampling the phase function}

After scattering, a new direction of the photon package is determined by sampling random values for $\theta$ and $\varphi$ from the phase function $\Phi(\theta,\varphi)$.
To accomplish this, we use the conditional probability technique. We reduce the phase function to the marginal distribution $\Phi(\theta)$,
\begin{equation}
\Phi(\theta)
=\int_0^{2\pi} \Phi(\theta,\varphi)\,\txd\varphi
=2\pi\ N\,S_{11}
=\frac{S_{11}}{\int_0^\pi S_{11}\sin\theta'\, \txd\theta'} \ .
\end{equation}
We sample a random $\theta$ value from this distribution through numerical inversion, that is to say, by solving the equation
\begin{equation}\label{eq:numInvTheta}
{\cal{X}} =\frac{\int_0^\theta S_{11}\sin\theta'\,\txd\theta'}{\int_0^\pi S_{11}\sin\theta'\, \txd\theta'}
\end{equation}
for $\theta$, where ${\cal{X}}$ is a uniform deviate, that is,\ a random number between $0$ and $1$ with uniform distribution.

Once we have selected a random scattering angle $\theta$, we sample a random azimuthal angle $\varphi$ from the normalized conditional distribution,
\begin{align}
\Phi_\theta(\varphi)
&=\frac{\Phi(\theta,\varphi)}{\int_0^{2\pi} \Phi(\theta,\varphi')\,\txd\varphi'}\\
&=\frac{1}{2\pi}\left(1+ P_{\text{L}}\,\frac{S_{12}}{S_{11}}\cos 2(\varphi - \gamma)\right),
\label{eq:varphiDistrib}
\end{align}
where the ratio $S_{12}/S_{11}$ depends on $\theta$.
This can again be done through numerical inversion, by solving the equation
\begin{align}
{\cal{X}}
&=\int_{0}^{\varphi}\Phi_{\theta}(\varphi')\,\txd\varphi' \\
&=\frac{1}{2\pi} \left( \varphi + P_{\text{L}}\,\frac{S_{12}}{S_{11}} \sin\varphi \cos(\varphi - 2\gamma)\right)
\label{eq:varphiSampling}
\end{align}
for $\varphi$, with ${\cal{X}}$ being a new uniform deviate.

\subsection{Updating the photon package}\label{RotationMethod}

After randomly selecting both angles $\theta$ and $\varphi$, we can use Eq.~\eqref{eq:genScatter} to adjust the main photon package's Stokes vector.
We can also calculate the outgoing propagation direction $\bfknew$ and the new reference normal $\bfnnew$ from the incoming propagation direction $\bfk$ and the previous reference normal $\bfn$ (see Fig.~\ref{PlaneRotation.fig}).
We use \textit{\textup{Euler's finite rotation formula}} \citep{Cheng1989} to rotate a vector $\bfv$ about an arbitrary axis $\vec{a}$ over a given angle $\beta$ (clockwise while looking along $\vec{a}$),
\begin{equation}\label{RodriguesRotation}
\bfvnew = \bfv \cos\beta+ (\vec{a} \times \bfv)\sin\beta+\vec{a}(\vec{a} \cdot \bfv)(1-\cos\beta).
\end{equation}
The last term of the right-hand side vanishes when the vector $\bfv$ is perpendicular to the rotation axis $\vec{a}$.

In our configuration, the reference normal $\bfn$ rotates about the incoming propagation direction $\bfk$ over the azimuthal angle $\varphi$. Because $\bfn$ is perpendicular to $\bfk$, we have
\begin{equation}
\bfnnew = \bfn \cos\varphi+ (\bfk \times \bfn)\sin\varphi.
\end{equation}
Furthermore, the propagation direction rotates in the current scattering plane, that is, about $\bfnnew$, over the scattering angle $\theta$. Again, $\bfk$ is perpendicular to $\bfnnew$, so that
\begin{equation}
\bfknew = \bfk \cos\theta+ (\bfnnew \times \bfk)\sin\theta.
\end{equation}

\subsection{Peel-off photon package}\label{sec:peeloff}

As described in Sect.~\ref{sec:SKIRT}, a common MCRT optimization is to send a peel-off photon package toward every observer from each scattering site.
The peel-off photon package must carry the polarization state after the peel-off scattering event, and its luminosity must be weighted by the probability that a scattering event would indeed send the outgoing photon package toward the observer.
To obtain this information, we need to calculate the scattering angles $\theta_\text{obs}$ and $\varphi_\text{obs}$, given the outgoing direction of the peel-off scattering event, or in other
words,\ the direction toward the observer, $\bfk_\text{obs}$.
This is effectively the scattering problem in reverse, in which random angles were chosen based on their probability, and the new propagation direction was calculated from these angles.

Finally, when the peel-off photon package reaches the observer, its Stokes reference direction must be rotated so that it lines up with the direction of north in the observer frame, $\bfk_\text{N}$, according to the \citet{IAU1974} conventions.
The scattering angle $\theta_\text{obs}$ is easily found through the scalar product of the incoming and outgoing directions,
\begin{equation}
\cos\theta_\text{obs} = \bfk \cdot \bfk_\text{obs}.
\end{equation}
Because $0\le\theta_\text{obs}\le\pi$, the cosine unambiguously determines the angle.

To derive the azimuthal angle $\varphi_\text{obs}$, we recall (Fig.~\ref{PlaneRotation.fig}) that it is the angle between the normal to the previous scattering plane, $\bfn$, and the normal to the peel-off scattering plane, $\bfn_\text{obs}$. The latter can be obtained through the cross product of the incoming and outgoing directions,
\begin{equation}
\bfn_\text{obs} = \frac{\bfk \times \bfk_\text{obs}} {||\bfk \times \bfk_\text{obs}||}.
\label{normal2}
\end{equation}
We need both cosine and sine to unambiguously determine $\varphi_\text{obs}$ in its $2 \pi$ range.
We easily have
\begin{equation}
\cos\varphi_\text{obs} = \bfn \cdot \bfn_\text{obs}.
\end{equation}
Because $\bfk$ is perpendicular to both $\bfn$ and $\bfn_\text{obs}$, the following relation holds,
\begin{equation}
\sin\varphi_\text{obs} \,\bfk= \bfn \times \bfn_\text{obs},
\end{equation}
or, after projecting both sides of the equation on $\bfk$,
\begin{equation}
\sin\varphi_\text{obs} = (\bfn \times \bfn_\text{obs}) \cdot \bfk.
\end{equation}
This derivation of $\varphi_\text{obs}$ breaks down for a photon package that happens to be lined up with the direction toward the observer before the peel-off event.
Indeed, in this special case of perfect forward or backward peel-off scattering, Eq.~\eqref{normal2} is undefined.
However, because the scattering plane does not change, it is justified to simply set $\varphi_\text{obs}=0$ instead.

We insert the calculated $\theta_\text{obs}$ and $\varphi_\text{obs}$ values into Eq.~\eqref{eq:genScatter} to adjust the peel-off photon package’s Stokes vector, and we also update the reference normal to $\bfn_\text{obs}$.
When the photon package's polarization state is recorded at the observer, its Stokes vector reference direction must be parallel to the north direction, $\bfk_\text{N}$, in the observer frame.
This is equivalent to orienting the reference normal along the east direction, $\bfk_\text{E}=\bfk_\text{obs} \times \bfk_\text{N}$.
The angle, $\alpha_\text{obs}$, between $\bfn_\text{obs}$ and $\bfk_\text{E}$ can be determined using a similar reasoning as for $\varphi_\text{obs}$, so that
\begin{equation}
\cos\alpha_\text{obs} = \bfn_\text{obs} \cdot \bfk_\text{E}
\end{equation}
and
\begin{equation}
\sin\alpha_\text{obs} = (\bfn_\text{obs} \times \bfk_\text{E}) \cdot \bfk_\text{obs}
.\end{equation}
The final adjustment to the Stokes vector is thus a rotation (see Eqs.~\ref{eq:applyRotation} and \ref{eq:rotation}) with the matrix $\bf{R}(\alpha_\text{obs})$. The Stokes vector is indifferent to rotations by $\pi$. Using $\bfk_\text{W}=-\bfk_\text{E}$ yields the same result.

\subsection{Reference direction for new photon packages}\label{sec:generateReferenceDirection}

We now return to the issue of selecting a reference direction, or more precisely, a reference normal, for newly emitted photon packages.
We stated at the end of Sect.~\ref{sec:referencedirection} that we can pick any direction perpendicular to the propagation direction, because we assume our sources to emit unpolarized radiation.
Indeed, it is easily seen from Eq.~\eqref{eq:varphiDistrib} that the probability distribution for the azimuthal angle $\varphi$ becomes uniform for unpolarized incoming radiation, that is,\ with $P_{\text{L,in}} = 0$.
Consequently, our choice of reference normal in the plane perpendicular to the propagation direction will be completely randomized after the application of the scattering transformation (Eq.~\eqref{eq:genScatter}).

We determine the reference normal, $\bfn = (n_{x},n_{y},n_{z})$, perpendicular to the propagation direction, $\bfk = (k_{x},k_{y},k_{z})$, using
\begin{align}
n_{x} &= -k_{x}k_{z} \, \bigg/ \, \sqrt[]{1-k_z^{2}}\\
n_{y} &= -k_{y}k_{z} \, \bigg/ \, \sqrt[]{1-k_z^{2}}\\
n_{z} &= \sqrt[]{1-k_z^{2}}.
\end{align}
When $\bfk$ is very closely aligned with the $Z$-axis, the root in these equations vanishes, and we instead select $\bfn = (1,0,0)$ as the reference direction.


\section{Validation}\label{sec:Validation}

\begin{figure}
\centering
\includegraphics[width = 0.9 \columnwidth, trim={7.5cm 1.7cm 7.6cm 0.8cm},clip]{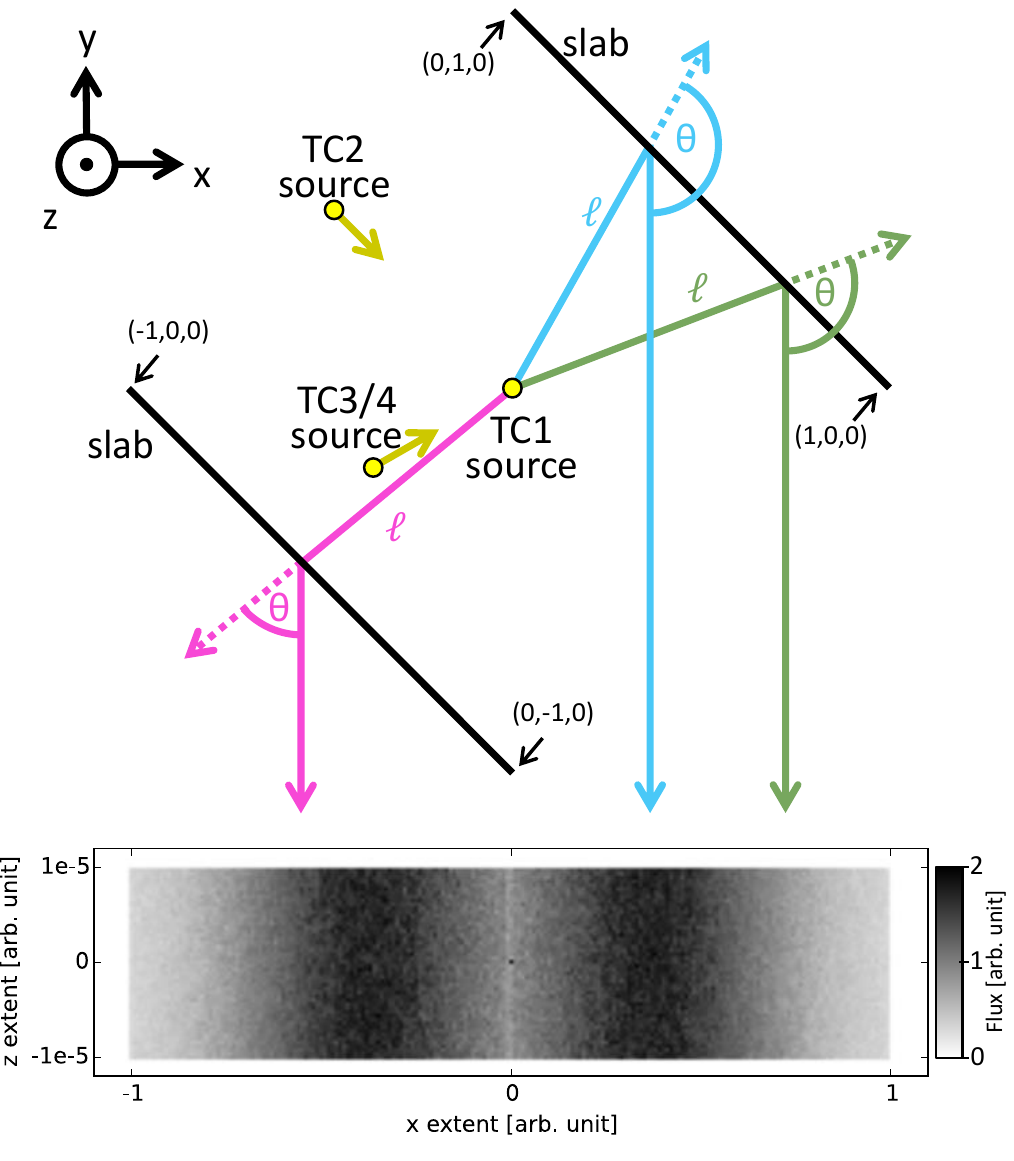}
\caption{\emph{Top}: Geometry used in the analytical test cases. The $z$-axis is toward the reader.
For test case 1, a central point source illuminates thin slabs of electrons that are slanted by $45\degree$ toward the observer.
The central source is replaced by a small blob of electrons for the other test cases. The blob is illuminated by a collimated beam from within the xy plane for test case 2 and from below the xy plane for test cases 3 and 4. \emph{Bottom}: Intensity map of test case 1 as seen from the observer.}
\label{PolarizationTest.fig}
\end{figure}

\begin{figure*}
\centering
\includegraphics[width = 1.9 \columnwidth]{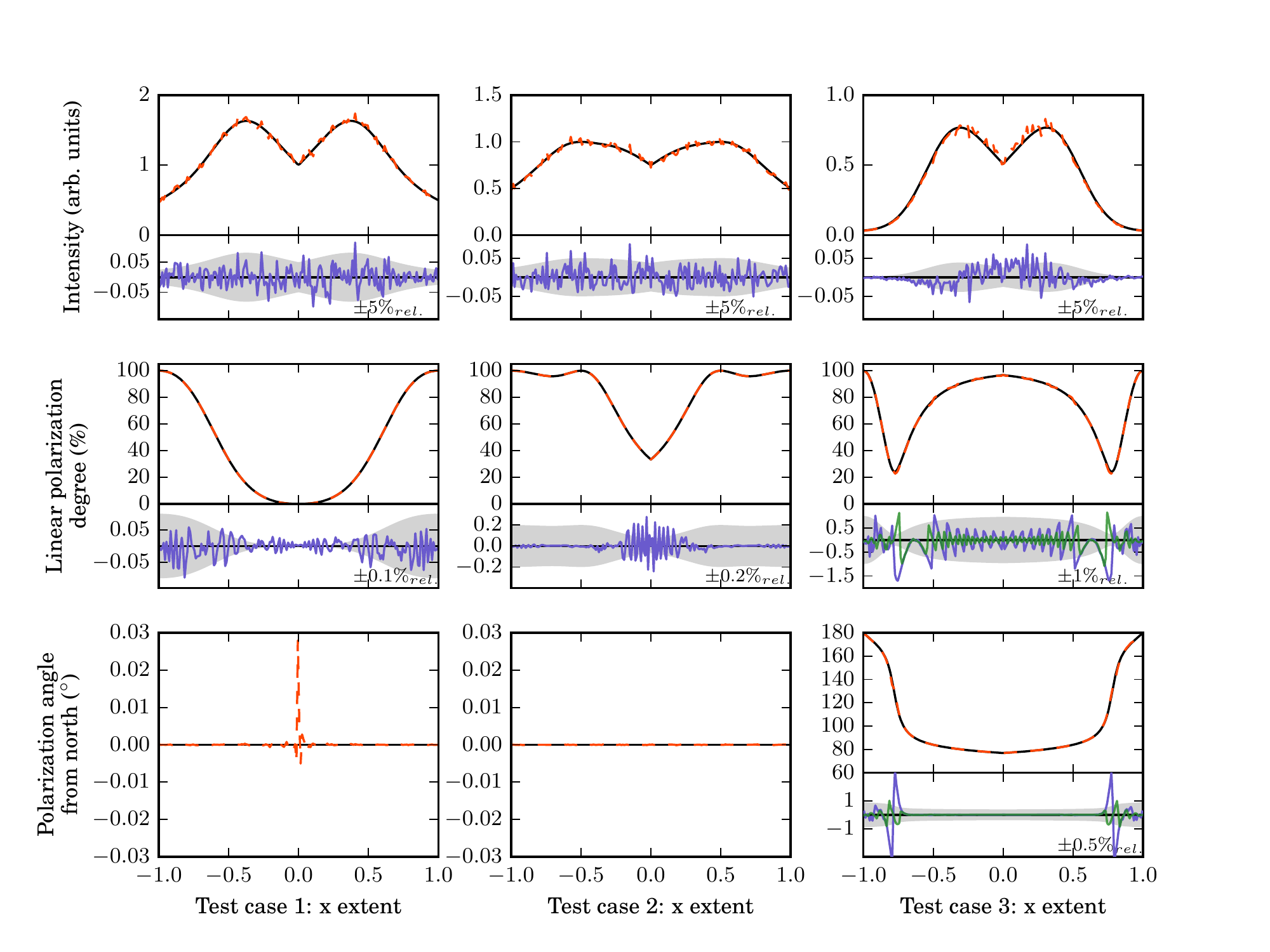}
\caption{Intensity (top row), linear polarization degree (middle row), and polarization angle (bottom row) of the observed radiation in test cases 1 through 3 (left to right columns).
The top section in each panel shows the analytical solution (black) and the model results (dashed orange).
The bottom section in each panel shows the absolute differences (blue) and relative differences (shaded area) of the analytic solution and the model. The green lines are calculated using twice the resolution of the $\theta$ scattering angle.}
\label{analyticalTestcases.fig}
\end{figure*}

\begin{figure}
\centering
\includegraphics[width = 0.95 \columnwidth]{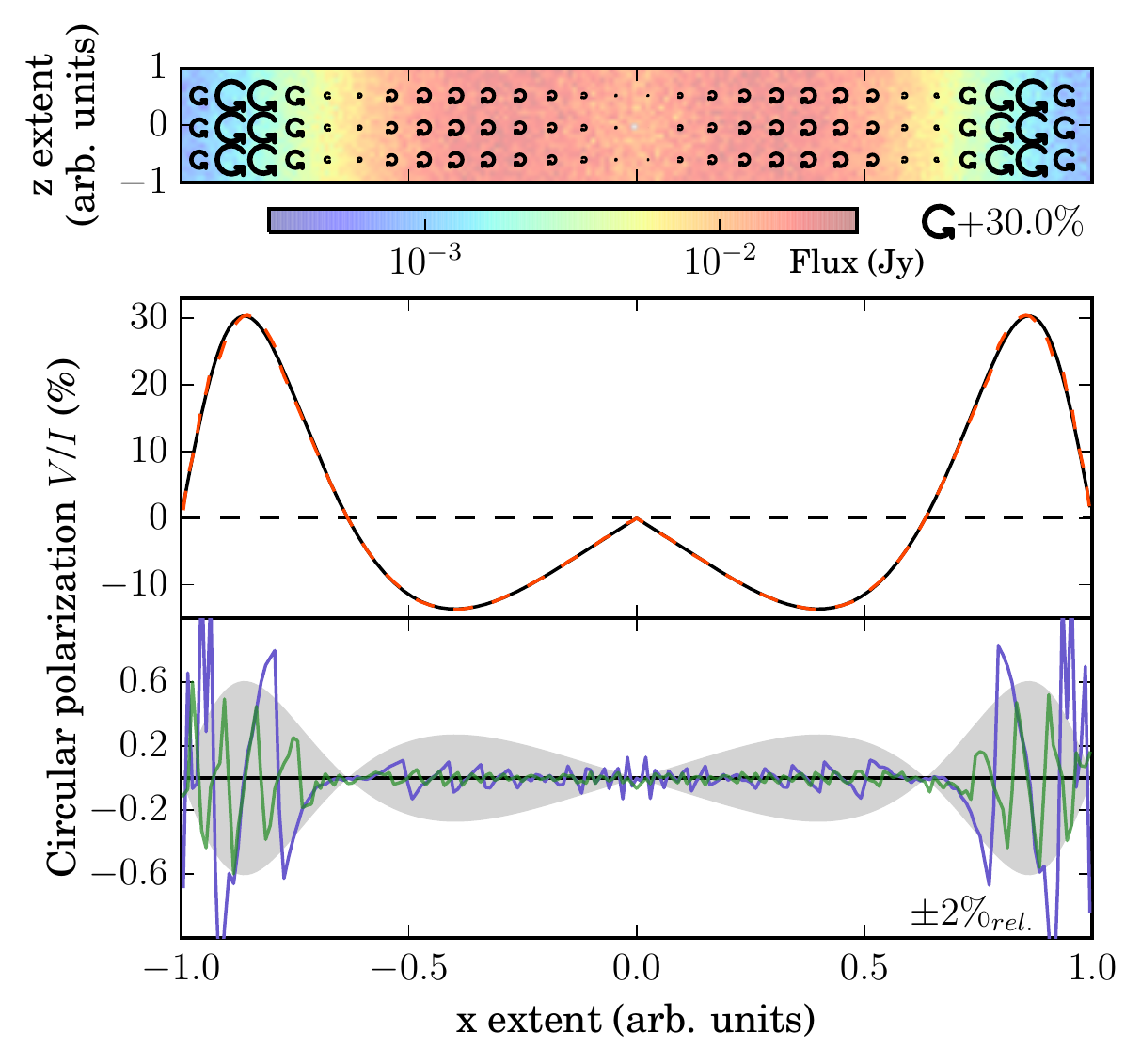}
\caption{Fraction of circular polarization, $V/I$, of the observed radiation in test case 4. The top panel shows the observed surface brightness overlaid with arrows indicating the handedness. The arrow length scales linearly with the magnitude.
The middle and bottom panels compare the analytical solution with the model results, as in Fig.~\ref{analyticalTestcases.fig}.}
\label{TC4plot.fig}
\end{figure}

\subsection{Test setup}\label{sec:testsetup}

In order to confirm the validity of our method and its implementation in SKIRT, we develop four test cases for which the results can be calculated analytically.
The analytical results are obtained solely using the formalisms of Sect.~\ref{sec:Polarization}, so that taken together, the test cases verify most aspects of the procedures presented in Sect.~\ref{sec:Method}.

The overall setup for the test cases is illustrated in Fig.~\ref{PolarizationTest.fig}.
A central source illuminates two physically and optically thin slabs of material, which scatter part of the radiation to a distant observer.
The slabs are arranged on the sides of a square rotated by $45\degree$ relative to the line of sight and are spatially resolved by the observer's instrument.
To simplify the calculations we only consider radiation detected close to the $xy$ plane, essentially reducing the geometry to two dimensions.
Because of the low scattering probability in the slabs, the path with the least number of scattering events will greatly dominate the polarization state at each instrument position.
This allows us to deduce the dominating scattering angle, $\theta$, corresponding to each instrument position along the $x$-axis.
Referring to Fig.~\ref{PolarizationTest.fig}, geometrical reasoning leads to
\begin{equation}\label{eq:testcasetheta}
\theta = \frac{\pi}{2} \pm \arctan\left(\frac{1}{{|x|}}-1\right)
,\end{equation}
with the plus sign for $x>0$ and the minus sign for $x<0$.
In combination with analytical scattering properties for the slab material, it then becomes possible to derive a closed-form expression for the components of the Stokes vector at each position.

Because the observer is considered to be at `infinite' distance, we can use parallel projection, and the distance from the slabs to the observer does not vary with $x$.
However, we do need to take into account the variations in the path length $\ell$ from the central source to the slabs because it affects the amount of radiation arriving at the slabs as a function of $x$.
Geometrical reasoning in Fig.~\ref{PolarizationTest.fig} again leads to
\begin{equation}\label{eq:pathlength}
\ell=\sqrt{x^2+(1-|x|)^2}.
\end{equation}

The scattering properties of the slabs and the makeup of the central source vary between the test cases.
For test cases 1 through 3, the slabs contain electrons, with scattering matrix ${\bf{M}}_\mathrm{Th}$ given by Eq.~\eqref{eq:ThomsonScattering}.
We study the observed intensity, $I$, the degree of linear polarization, $P_{\text{L}}$, and the linear polarization angle, $\gamma$, of these test cases in Sect.~\ref{sec:testcases123}.
Scattering by electrons never causes circular polarization.
Therefore in Sect.~\ref{sec:testcase4} we introduce slabs that contain synthetic particles with custom-designed scattering properties (test case 4). This allows us to study the observed circular polarization.

Test case 1 has a central point source. For the remaining test cases, the central source is replaced by a small blob of electrons illuminated by a collimated beam positioned at varying angles, so that the center becomes the site of first scattering.

A numerical implementation of the test cases will always discretize certain aspects of the theoretical test setup.
For our implementation in SKIRT, we made the following choices.
The spatial domain of the setup is divided into $601\times601\times61$ cuboidal grid cells lined up with the coordinate axes (we use odd numbers to ensure that the origin lies in the center of a cell).
The cells overlapping the slabs (and where applicable, the central blob) contain electrons, the other cells represent empty space.
The edges of the cells and slabs are not aligned. Each detector pixel provides averages over multiple cells, and the slabs are optically thin ($\tau=10^{-3}$ along their depth). The length of the slabs is 1.141, their depth is 0.006, and their height is $2\times10^{-5}$.
The detector in the observer plane has a resolution of 201 pixels along the $x$-axis.
We use~$10^{10}$ photon packages for each test case to minimize the stochastic noise characteristic of Monte Carlo codes.
SKIRT uses precomputed tables of various quantities with a resolution of $1\degree$ in $\theta$ and $\varphi$, to help perform the numerical inversions in Eqs.~\eqref{eq:numInvTheta} and \eqref{eq:varphiSampling}.

\subsection{Linear polarization}\label{sec:testcases123}

Test case 1 is designed to test the peel-off procedure described in Sect.~\ref{sec:peeloff}.
The slabs contain electrons, and the central point source emits unpolarized photon packages.
When a photon package's direction is toward one of the slabs, the forced interaction algorithm of SKIRT initiates a scattering event at the slab and a peel-off photon package is sent toward the observer.
Because the slabs are optically thin, the luminosity of the scattered original photon package
is small. It is subsequently deleted and a new photon package is launched.

As a result, the Stokes vector of the observed photon packages can be written as
\begin{equation}
{\bfS^{\text{TC1}}}=\ell^{-2}{\bf{R}}\left(\frac\pi2\right)\ {\bf{M}}_\mathrm{Th}(\theta)\ (1,0,0,0)^\text{T},
\end{equation}
reflecting from right to left a scattering transformation starting from an unpolarized state (and thus $\varphi=0$), a rotation to align the reference direction with the observer frame, and the dependency on the path length from the source to the slab.
We can now substitute Eqs.~\eqref{eq:rotation}, \eqref{eq:ThomsonScattering}, \eqref{eq:testcasetheta}, and \eqref{eq:pathlength} into this equation.
Because the arctangent of  Eq.~\eqref{eq:testcasetheta} is used as an argument for the cosine in Eq.~\eqref{eq:ThomsonScattering}, the final equations for the intensity and the linear polarization degree reduce to polynomials,
\begin{align}
I^\text{TC1} &= \frac{3x^2-4|x|+2}{2(2x^2-2|x|+1)^2},\\
P_\text{L}^\text{TC1} &= \frac{x^2}{3x^2-4|x|+2}.\label{eq:TC1_PL}
\end{align}
The orientation of the polarization is perpendicular to the scattering plane, that is, north/south or along the $y$-axis in the observer frame.

In test case 2 we add a scattering event to verify part of the procedures for scattering the main photon package.
To this end, we replace the central point source with a small blob of electrons at the same location, and illuminate this electron blob with a collimated beam positioned at $(-1,1,0)$ and oriented parallel to the slabs toward the bottom right (see Fig.~\ref{PolarizationTest.fig}).
In this setup, a photon package emitted by the collimated source can reach the slabs only after being scattered by the electrons in the central blob.
This effectively adds a forced first scattering event to all photon packages reaching the observer, with a scattering angle that can be deduced from the geometry.
The peel-off scattering angle is still given by Eq. \eqref{eq:testcasetheta}.
The scattering angle for the first scattering event in the central electron blob is
\begin{equation}
\theta_1 =\theta \pm \left(-\frac\pi4\right),
\end{equation}
again with the plus sign for $x>0$ and the minus sign for $x<0$.
Because all components of the setup are in the same plane, the scattering plane is always the same (the $xy$-plane).
The Stokes vector of the observed photon packages can thus be written as\begin{equation}
{\bfS^{\text{TC2}}}=\ell^{-2}{\bf{R}}\left(\frac\pi2\right)\ {\bf{M}}_\mathrm{Th}(\theta)\ {\bf{M}}_\mathrm{Th}(\theta_1)\ (1,0,0,0)^\text{T}.
\end{equation}
The intensity and linear polarization degree for test case 2 are
\begin{align}
I^\text{TC2} &= \frac{12x^4-28|x|^3+29x^2-14|x|+3}{4(2x^2-2|x|+1)^3},\\
P_\text{L}^\text{TC2}  &= \frac{4x^4-4|x|^3+3x^2-2|x|+1}{12x^4-28|x|^3+29x^2-14|x|+3}.
\end{align}
The orientation of the polarization is north/south.

In the third test case we move the collimated source below the $xy$-plane to $(-\sqrt3,-1,-4-2\sqrt3)$, so that we can test the rotation of the Stokes vector reference direction between scattering events.
The source is now placed at an inclination of $165\degree$ (relative to the $z$-axis) and an azimuthal angle of $30\degree$ (clockwise from the $x$-axis).
It still points toward the central electron blob, that is,\ toward the top right and out of the page in Fig.~\ref{PolarizationTest.fig}.
As a result, the normal of the first scattering plane is tilted, while the normal of the peel-off scattering plane remains aligned with the $z$-axis.
With some trigonometry, we arrive at expressions for the angles involved in the first (main) and second (peel-off) scattering events,
\begin{align}
\theta_1 &= \arccos\left(\pm\frac{-1+\sqrt{3}+(4-2\sqrt{3})|x|}{4\sqrt{2-4|x|+4x^2}}\right), \label{eq:testcase3theta1}\\
\varphi_1 &= \pm \arctan\left(\frac{2(1+\sqrt{3})\sqrt{1-2|x|+2x^2}}{3-\sqrt{3}-2|x|}\right), \label{eq:testcase3phi1} \\
\theta_2 &= \pi/2 \pm \arctan(1/|x| - 1), \label{eq:testcase3theta2} \\
\varphi_2 &= \pi/2, \label{eq:testcase3phi2}
\end{align}
with the plus sign for $x>0$ and the minus sign for $x<0$.
The expression provided in Eq.~\eqref{eq:testcase3phi1} for $\varphi_1$ is simplified and shifted by~$\pm\pi$ for some $x$.
The Stokes vector is invariant under rotations by $\pi$.

The Stokes vector of the observed photon packages can now be written as
\begin{equation}\label{eq:testcase3Stokes}
{\bfS^{\text{TC3}}}=\ell^{-2}{\bf{R}}(\varphi_2)\ {\bf{M}}_\mathrm{Th}(\theta_2)\ {\bf{R}}(\varphi_1)\ {\bf{M}}_\mathrm{Th}(\theta_1)\ (1,0,0,0)^\text{T}.
\end{equation}
To limit the complexity of presentation, we provide expressions for the Stokes parameters from which the linear polarization degree and angle can be calculated using Eqs.~\eqref{eq:degreeOfLinearPolarisation} and~\eqref{eq:directionOfPolarisation},
\begin{align}
\begin{split}
I^\text{TC3} =& \frac1{32\ell^6}\left[(62-16\sqrt{3})x^4-(150-30\sqrt{3})|x|^3\right.\\
&             \left.+(156-25\sqrt{3})x^2-(78-8\sqrt{3})|x|+18-\sqrt{3}\right],
\end{split}\\
\begin{split}
Q^\text{TC3} =& \frac1{32\ell^6}\left[(2-16\sqrt{3})x^4+(22+34\sqrt{3})|x|^3\right.\\
&   \hspace{-1mm}  \left.-(28+39\sqrt{3})x^2+(14+24\sqrt{3})|x|-(2+7\sqrt{3})\right],
\end{split}\\
U^\text{TC3} =& \frac1{8\ell^4}\left[(1+\sqrt{3})x^2-(1+2\sqrt{3})|x|+\sqrt{3}\right],\\
V^\text{TC3} =& \ 0.
\end{align}

Figure~\ref{analyticalTestcases.fig} compares the analytical solutions and SKIRT results for the observed intensity, linear polarization degree, and polarization angle for these three test cases.
The intensity curves show a relative noise level of on average about 3\%. The linear polarization degrees are identical to below 0.1\% absolute for test cases 1 and 2 and below 1\% absolute for test case 3. The polarization angles from north are correct to below $0.05^\circ$ for test cases 1 and 2 and below $1^\circ$ for test case 3.
While the linear polarization degree and position angle can be determined from a simulation with relatively few photon packages, reducing the noise in the intensity requires significantly more photon packages. This is because the number of photon packages arriving at each pixel is subject to Poisson noise, whereas the path that each photon package takes to the same pixel is defined within tight boundaries. The intensity curve depends on the number of photons. The linear polarization degree and angle are independent of the number of photon packages. Their noise is due to the size of the pixels. Slightly different paths and scattering angles might contribute to the same pixel.

The polarization angle for test case 1 shows an intriguing spike near $x=0$.
As we can see in the linear polarization degree curve and from Eq.~\eqref{eq:TC1_PL}, the radiation arriving at $x=0$ is unpolarized. This implies that both the $Q$ and $U$ components of the Stokes vector are zero and the polarization angle becomes undefined (see Eq.~\ref{eq:directionOfPolarisation}). This in turn causes small numerical inaccuracies in the calculations for photon packages arriving very close to $x=0$.

The relative differences for the other quantities are generally smaller than a fraction of a percent.
In the results of test case~3 the residuals of the linear polarization degree and polarization angle contain spikes that are resolved by multiple pixels each and symmetric with respect to $x=0$. The residual of the intensity curve shows a similar effect. It tends to be lower than expected in the outer regions and higher than expected in the inner region. These orderly deviations indicate a systematic difference rather than noise. In fact, these discrepancies are caused by resolution effects in the SKIRT implementation of our method (see Sect.~\ref{sec:testsetup} for a description of our discretization choices).
For example, consider the interval $0.6<x<0.7$ for test case~3, which is resolved by 10 pixels along the $x$-axis of the detector in the observer frame.
The corresponding interval for scattering angle $\theta_1$ (see Eq.~\ref{eq:testcase3theta1}) is $75.0\degree<\theta_1<75.1\degree$, that is, only a fraction of the $1\degree$ resolution in the SKIRT calculations related to $\theta$.
It is obvious that this lack of angular resolution relative to the output resolution will cause inaccuracies.
We calculated the residuals in the polarization degree and angle from north using twice the $\theta$ resolution and show them in pale green in Fig.~\ref{analyticalTestcases.fig}. They confirm that increasing the $\theta$ resolution in the calculations indeed reduces the discrepancies accordingly.

\subsection{Circular polarization}\label{sec:testcase4}

To include circular polarization in test case 4, we use synthetic particles similar to electrons, but with a scattering matrix that mixes the $U$ and $V$ components of the Stokes vector:
\begin{equation}\label{eq:synthElectrons}
{\bf{M}}_\mathrm{syn}(\theta)
=\frac12
\begin{pmatrix}
\cos^2\theta+1 & \cos^2\theta-1 & 0 & 0 \\
\cos^2\theta-1 & \cos^2\theta+1 & 0 & 0 \\
0 & 0 & 2\cos^2\theta & -2\cos\theta\sin\theta \\
0 & 0 & 2\cos\theta\sin\theta & 2\cos^2\theta
\end{pmatrix}.
\end{equation}
We use the same geometry as for test case 3, replacing the electrons in the two slabs and in the central blob by particles described by Eq.~\eqref{eq:synthElectrons}. The angles are still described by Eqs.~\eqref{eq:testcase3theta1} through \eqref{eq:testcase3phi2}, Eq.~\eqref{eq:testcase3Stokes} still holds, and the Stokes parameters of the observed photon packages become\begin{align}
I^\text{TC4} =& I^\text{TC3}, \\
Q^\text{TC4} =& Q^\text{TC3}, \\
\begin{split}
U^\text{TC4} =& \frac{\pm 1}{8\ell^5}\left[(1+\sqrt{3})|x|^3-(2+3\sqrt{3})x^2\right.\\
             & \hspace{20mm} \left.+(1+3\sqrt{3})|x|-\sqrt{3}\right] \label{eq:testcase4U},
\end{split}\\
V^\text{TC4} =& \frac{1}{8\ell^5}\left[-(1+\sqrt{3})|x|^3+(1+2\sqrt{3})x^2-\sqrt{3}|x|\right].
\end{align}
In Eq.~\eqref{eq:testcase4U} the plus sign is again for $x>0$ and the minus sign for $x<0$.

Figure~\ref{TC4plot.fig} shows the (relative) circular polarization, $V/I$, for this test case, again comparing the analytical solutions with the SKIRT results.
The relative differences between the analytical and simulated results for $|x|<0.7$ are below one percent.
The larger discrepancies for $|x|>0.7$ are again due to the limited resolution in the SKIRT calculations related to $\theta$. The pale green line again shows the residuals when calculating with $0.5\degree$ resolution and has a significantly smaller residual curve. Disregarding the outer part, the circular polarization is correct to 0.1\% absolute.


\section{Benchmark tests}\label{sec:Models}

\begin{figure*}
\centering
\includegraphics[width=1.6\columnwidth]{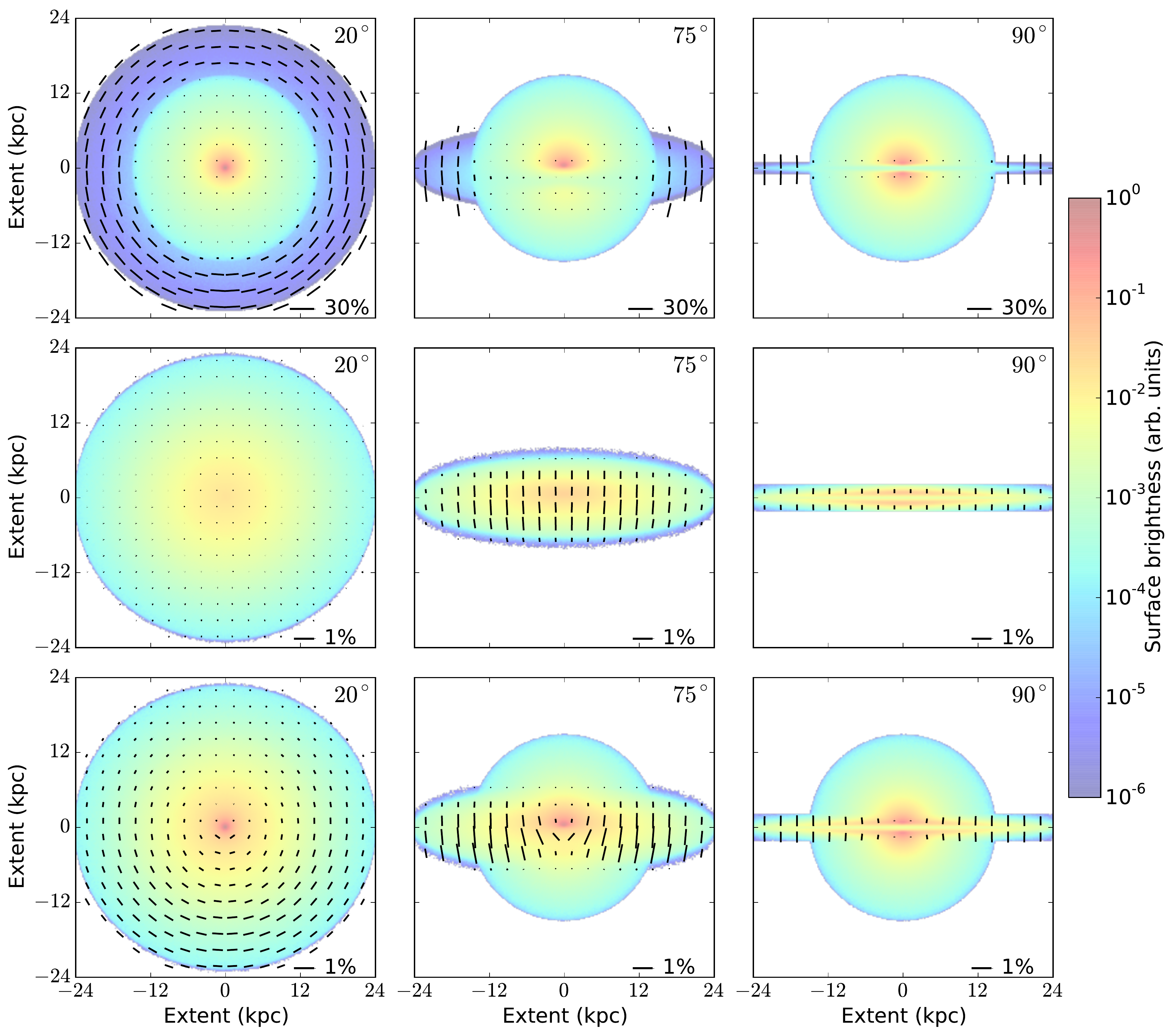}
\caption{$B$-band surface brightness maps (color scale) overlaid with linear polarization maps (line segments) for inclinations of 20\degree (left), 75\degree (middle), and 90\degree (right column). The top row shows a model with only a stellar bulge, the middle row a model with only a stellar disk, and the bottom row a model with both stellar bulge and disk with a ratio of bulge to total luminosity of B/T=0.5. The dust disk is the same in the three cases and has a central face-on $V$-band optical depth of 10.}
\label{OurBianchi.fig}
\end{figure*}

\begin{figure*}
\centering
\includegraphics[width = 1.9 \columnwidth]{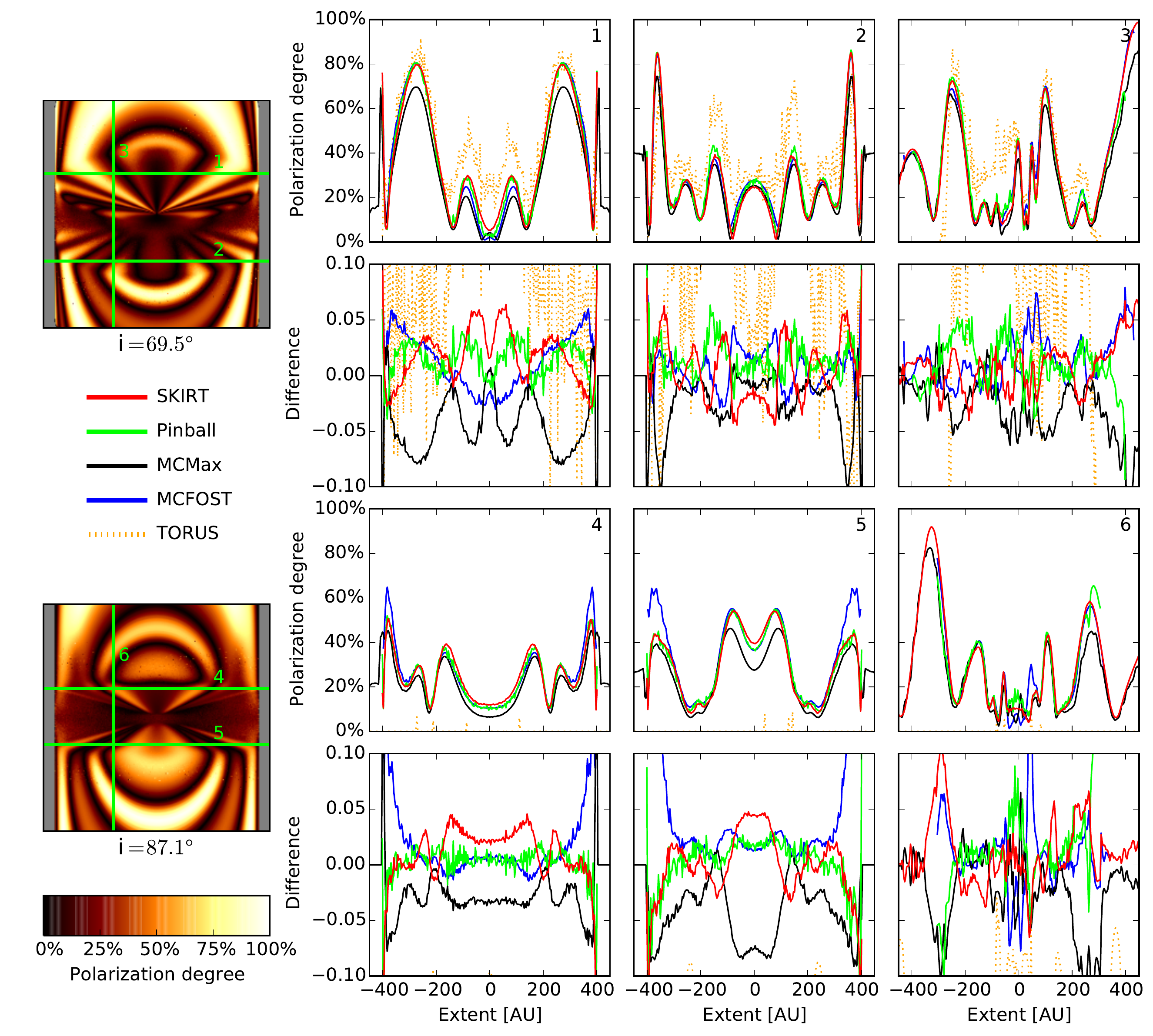}
\caption{Polarization degree maps for two inclinations of a disk with high optical density ($\tau = 10^6$). On the left are the linear polarization degree maps we calculated using SKIRT. The dust grain size is the same as the wavelength (\mum{1}), creating the intricate pattern of the polarization degree. On the right are cuts 1--6 through the maps along with results of various codes. Pinball \citep{Watson2001} in green, MCMax \citep{Min2009} in black, MCFOST \citep{Pinte2006} in blue, TORUS \citep{Harries2000} in dashed orange, and SKIRT in red. The data of the other codes were taken from the official website.}
\label{pinteBenchmark.fig}
\end{figure*}

\subsection {Disk galaxy}\label{sec:bianchi}

We compare results of our code to \citet{Bianchi1996} as a first test of our implication of dust scattering polarization (rather than just Thompson scattering). Bianchi and collaborators describe the polarization effects of scattering by spherical dust grains in monochromatic MCRT simulations of 2D galaxy models at the B band (\mum{0.44}) and I band (\mum{0.9}).
The models include a stellar bulge, a stellar disk, and a dust disk.
The stellar bulge is described by \citet{Jaffe1983} (scale radius \kpc{1.86}, truncated at \kpc{14.8}), and the stellar disk is a double-exponential disk (scale length \kpc{4}, truncated at \kpc{24}; scale height \kpc{0.35}, truncated at \kpc{2.1}).
The relative strength of the stellar components and the stellar sources is varied.
In all models, the dust is assumed to be homogeneously distributed in a double-exponential disk embedded in the stellar disk, with the same horizontal parameters as the stellar disk, but much thinner vertically (scale height \kpc{0.14}, truncated at \kpc{0.84}). The central face-on optical depth of the dust disk $\tau_\mathrm{V}$ is a free parameter.
The properties of the dust are taken from the MRN dust model \citep{Mathis1977}, which includes graphite grains and astronomical silicates with a size distribution $n(a)\propto a^{-3.5}$, $\mum{0.005}<a<\mum{0.25}$.
Polycyclic aromatic hydrocarbons (PAHs) and very small grains are not included.

Figure~\ref{OurBianchi.fig} displays our results, which correspond to the \citet{Bianchi1996} model shown in their Figure~8.
We use the same geometry, described above, and the same bulge-to-total ratio (B/T=0.5), wavelength ($\lambda=\mum{0.44}$), and optical depth ($\tau_\mathrm{V}=10$).
We use a different dust model. It includes a wider range of graphite and silicate grain sizes following the \citet{Zubko2004} size distribution, in addition to a small fraction of PAHs as described in \citet{Camps2015b}.
The differences between the dust models do not affect the qualitative results at optical wavelengths, but do prevent a quantitative comparison.

Surface brightness maps (color scale) overlaid with a linear polarization maps (line segments) are shown in Fig.~\ref{OurBianchi.fig}. The inclinations range from nearly face-on ($20^\circ$, left) to edge-on ($90^\circ$, right).
The top and middle rows show models from which the stellar disk and the stellar bulge, respectively, was removed. The bottom row shows the B/T=0.5 model, that is to say, the model including both stellar bulge and stellar disk.
The dust disk is identical for the three models.

Overall, our results are compatible with those reported by \citet{Bianchi1996}.
The largest polarization degrees are observed near the major axis.
For pure disks, the face-on view shows very little polarization. The polarization degree increases with inclination to a maximum near an inclination of about 80\degree\ and slightly decreases again when approaching $90\degree$. The polarization degree averaged over all pixels is below 1\% for all inclinations.
For pure bulges, the maximum polarization degree is largely independent of the inclination. In the outer regions of the dust disk, the linear polarization degree is 22\%.
In the mixed B/T=0.5 model, the polarization degrees are drastically reduced. We find values up to 1.6\%, comparable to the results of \citet{Bianchi1996}, who find 1 to 1.5\%.

We also test the corresponding model for the $I$ band ($\lambda=\mum{0.9}$) with a color-adjusted bulge luminosity ($B-I$ = 1 mag), again confirming overall agreement \citet{Bianchi1996}.

\subsection{Dusty disk around star}
To test the performance of our code on a problem of a different scale, we compute polarization results of \citet{Pinte2009}. In it, a thick dusty disk surrounds a central star. The star extends out to \unit[2]{AU} and has a temperature of \unit[4000]{K}. The dust consists of spherical grains with a radius of \mum{1}, and the light has a wavelength of \mum{1} as well. The dust density distribution $\rho$ is cylindrical,
\begin{equation}
\rho(R,z) = \frac{3\Sigma_0}{2R_0} \left(\frac{R}{R_0}\right)^{-5/2} \exp\left[-\frac12 \left(\frac{z}{h_0}\right)^2 \left(\frac{R}{R_0}\right)^{-9/4}\right]
,\end{equation}
with a surface density $\Sigma_0$, scale radius $R_0 = \AU{100}$, radial distance from the center $R$, vertical distance from the midplane $z,$ and scale height $h_0 = \AU{10}$. The disk is truncated at $R_{min} = \AU{0.1}$ and $R_{max} = \AU{400}$. The surface density depends on the total dust mass $m = \unit[3 \times 10^{-5}]{M_\odot}$ by
\begin{equation}
\Sigma_0=m/\left[\frac{12}{5} (2\pi)^{3/2}  h_0 R_0 \left((R_{min}/R_0)^{5/8}-(R_{max}/R_0)^{5/8}\right) \right]
\end{equation}
and the albedo of the dust is 0.6475 and the opacity \unit[4752]{cm$^2$/g} at \mum{1}. We adopt the scattering matrix as provided by \citet{Pinte2009}.\footnote{http://ipag.osug.fr/~pintec/benchmark/} The system is resolved at a distance of \unit[140]{pc} by a detector with $251 \times 251$ pixels covering $900 \times 900$~$\textrm{AU}^2$.

Figure~\ref{pinteBenchmark.fig} shows the linear polarization maps calculated by our code for inclinations of 69.5\degree{} and 87.1\degree. The flux difference from the borders to the center area is about 17 orders of magnitude. The maps are displayed in gray outside the truncation radius, where no flux was recorded. The intricate pattern of the polarization degree is a result of the uniform grain radius being equal to the wavelength. The phase function therefore contains resonances and steep gradients for small changes of the scattering angle. We compare our results to the results of the four polarization capable codes in \citet{Pinte2009} along six cuts through the maps. The first and third row show the polarization degree along the cuts, and the second and fourth row show the difference of the codes to the average of all results. The TORUS code \citep{Harries2000} is not included in calculating the averages because its signal-to-noise ratio is too low. In the central area the codes agree within 10\% of absolute polarization, but as the true result is unknown, a quantitative analysis of the results of this benchmark is difficult. In general, the results of the SKIRT code are close to the average result, and SKIRT seems to agree particularly well with the Pinball code \citep{Watson2001}. Pinball employs some of the same optimization techniques that SKIRT uses (e.g., forced interaction and peel-off, named ``forced escape'' in their paper).


\section{Application: spiral galaxy models}\label{sec:spiralarmmodel}

\begin{figure*}
\begin{center}
\includegraphics[width=1.8\columnwidth]{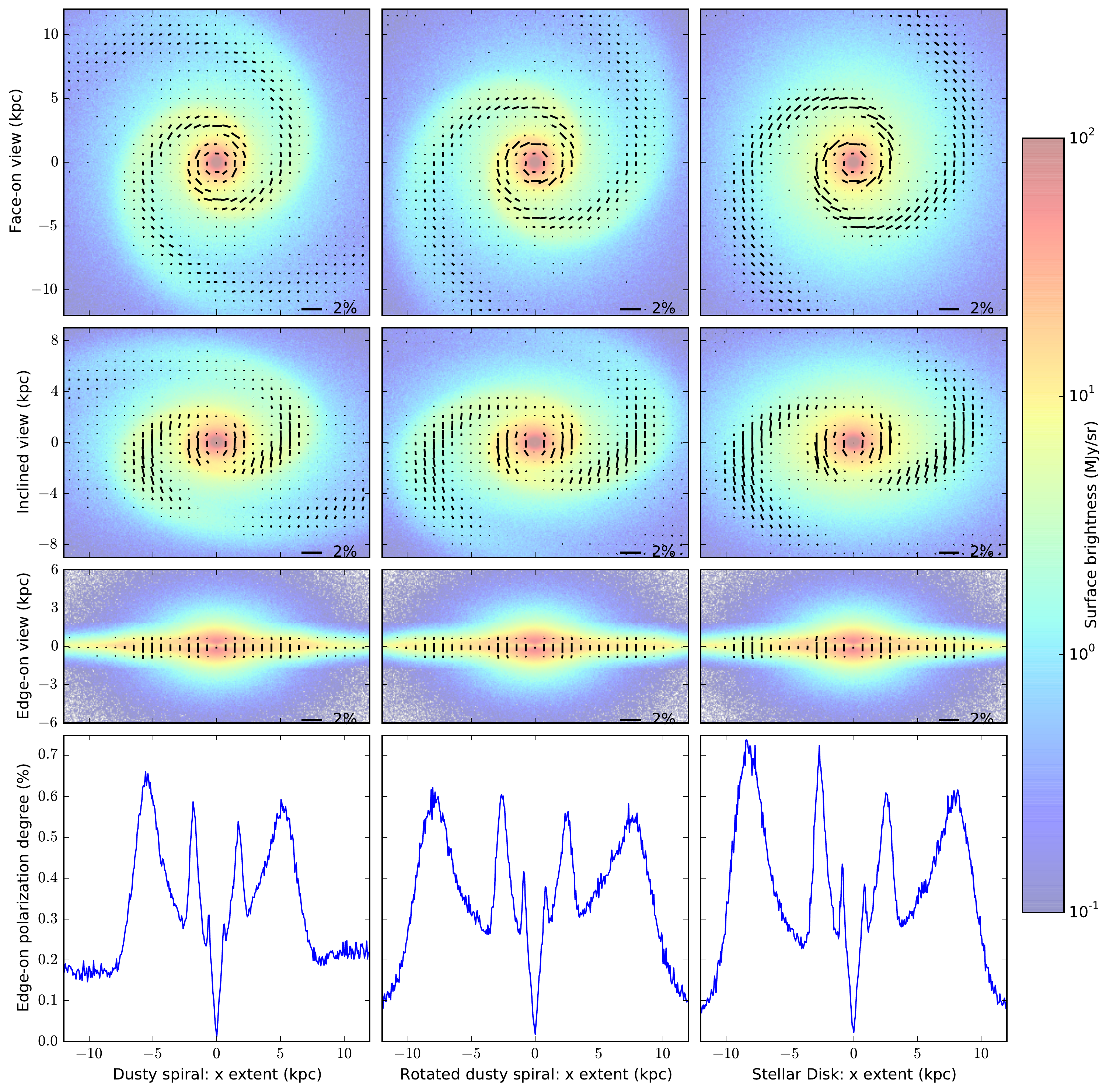}
\end{center}
\caption{Spiral galaxy model described in Sect.~\ref{sec:spiralarmmodel} and Table~\ref{tab:SpiralParams} observed at a wavelength of \mum{1}.
Rows from top to bottom: face-on, inclined ($53\degree$), and edge-on surface brightness (color scale) overlaid with linear polarization degree and orientation (line segments); linear polarization degree of the edge-on view, averaged over the vertical axis.
Columns from left to right: the reference model; the same model observed from a different azimuth angle (rotated clockwise by 120\degree); a model without the spiral arm perturbations in the stellar disks, with the rotated orientation.}
\label{fig:centralFig}
\end{figure*}

\begin{figure}
\begin{center}
\includegraphics[width= 0.95\columnwidth]{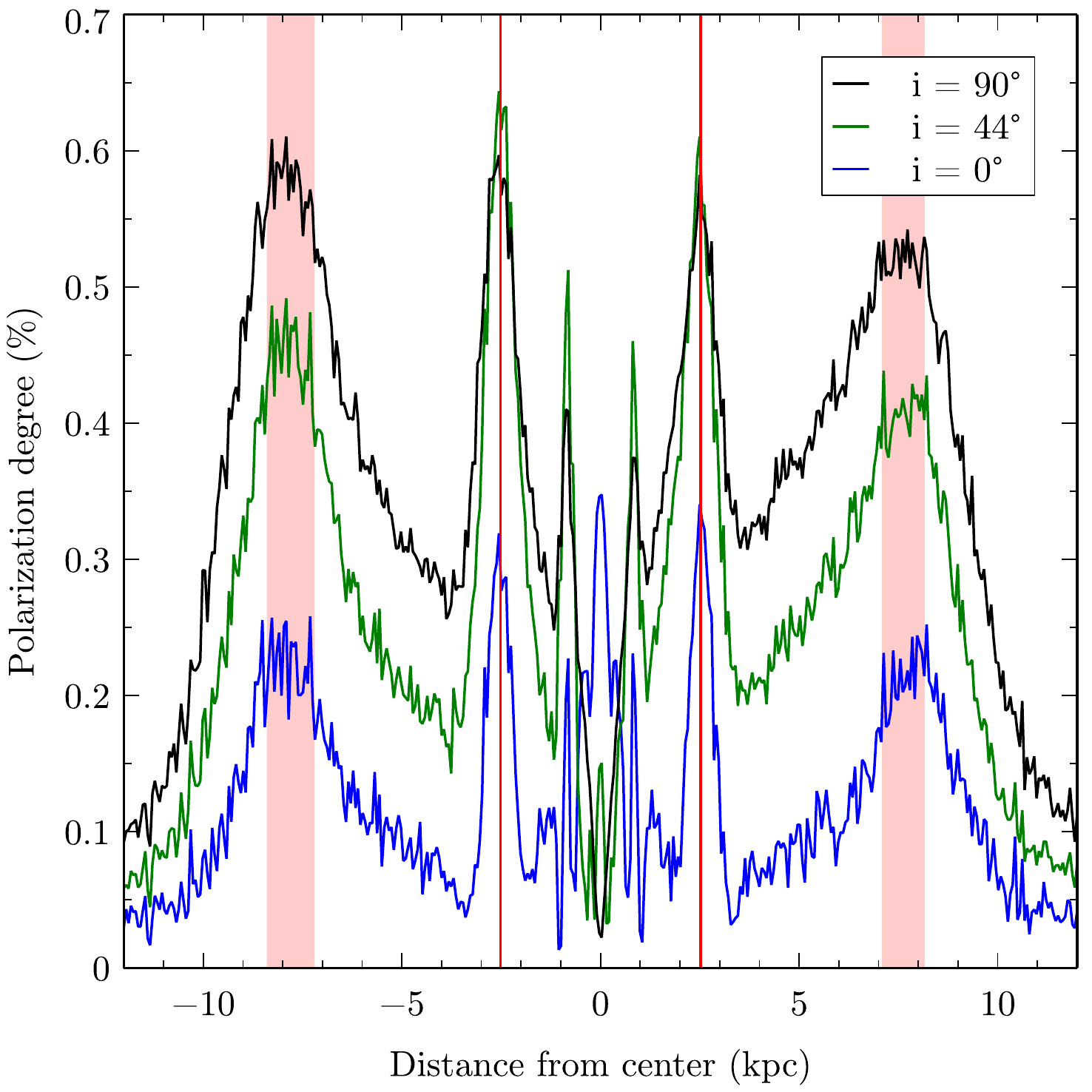}
\end{center}
\caption{Linear polarization degree, averaged over the vertical axis, of the rotated model (column B of Fig.~\ref{fig:centralFig}) observed at \mum{1}, for inclinations ranging from edge-on (i=90\degree) to face-on (i=0\degree).}
\label{fig:inclinations}
\end{figure}

\begin{figure}
\begin{center}
\includegraphics[width=0.95\columnwidth]{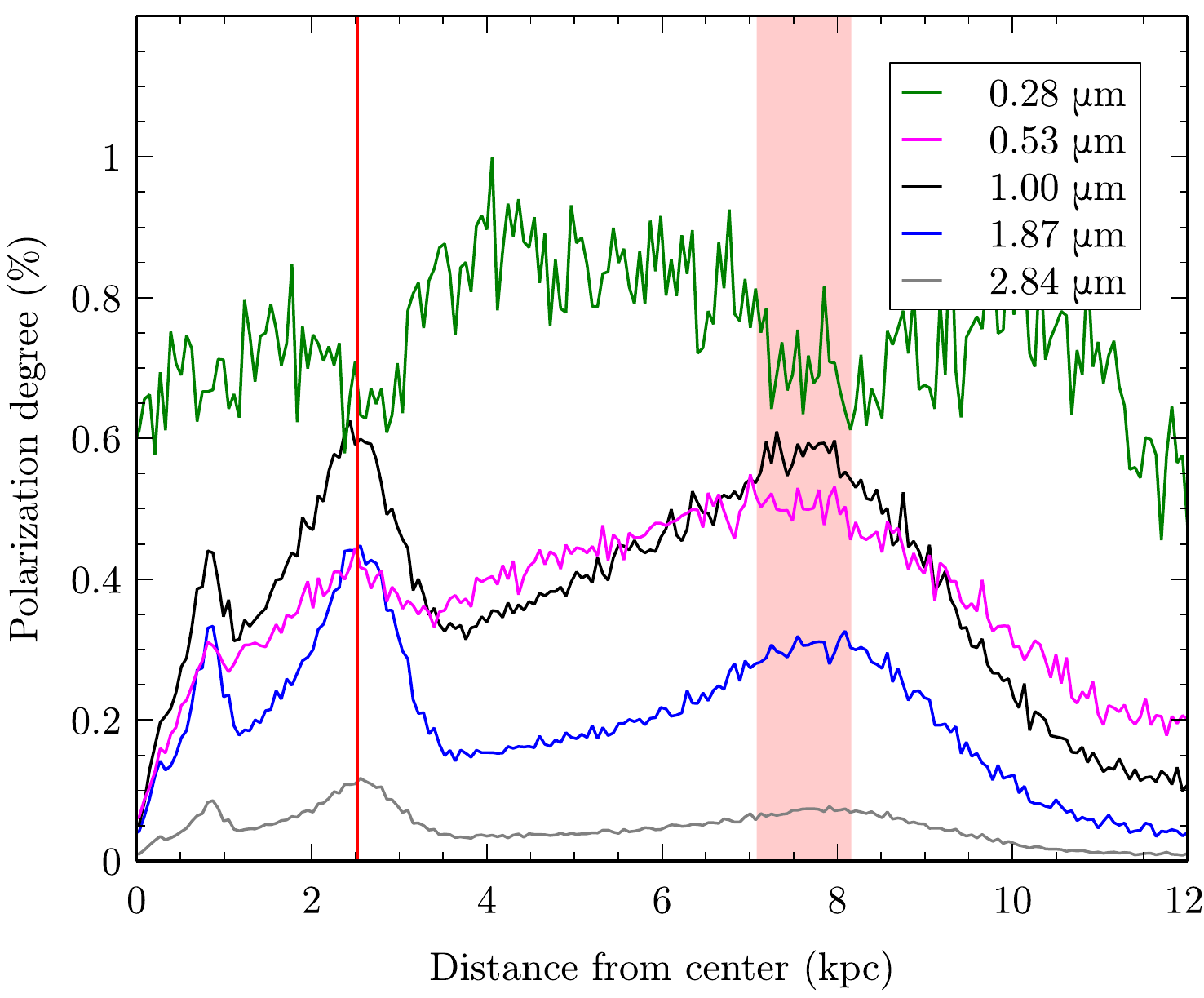}
\caption{Averaged linear polarization degree for one side of the edge-on view of the rotated model, observed at optical and near-infrared wavelengths from \mum{0.28} to \mum{2.84}. The red vertical bands correspond to the bands in Fig.~\ref{fig:inclinations} and trace the tangent points of the dust spiral arms.}
\label{fig:multWavelength}
\end{center}
\end{figure}

We study the polarization properties of a 3D galaxy model including spiral arms to investigate how the spiral arm structure is imprinted in the polarization structure. We also study this in the edge-on view when the structure is not easily characterized from the intensity alone.

\begin{table}
\centering
\caption{Parameters for our spiral galaxy model, including the blackbody temperature and the bolometric luminosity of the stellar components, and the total dust mass in the dust component.}
\label{tab:SpiralParams}
\def\arraystretch{1.1}
\setlength\tabcolsep{5pt}
\begin{tabular}{llll}
    \toprule
S\'ersic profile            & Bulge     \\
    \midrule
S\'ersic index          & 3         \\
Effective radius            & \kpc{1.6} \\
Flattening parameter        & 0.6           \\
Temperature         & 3500 K        \\
Luminosity          & $3\times10^{10}\,\mathrm{L}_\odot$        \\
    \bottomrule
    \toprule
Exponential disks       & Old stars     & Young stars   & Dust      \\
    \midrule
Scale length            & \kpc{4}       & \kpc{4}       & \kpc{4}       \\
Horizontal cutoff       & \kpc{20}      & \kpc{20}      & \kpc{20}      \\
Scale height            & \kpc{0.35}    & \kpc{0.2} & \kpc{0.2} \\
Vertical cutoff         & \kpc{1.75}    & \kpc{1}       & \kpc{1}       \\
Number of arms      & 2         & 2         & 2         \\
Pitch angle         & 20\degree & 20\degree & 20\degree \\
Radius zero-point       & \kpc{4}       & \kpc{4}       & \kpc{4}       \\
Phase zero-point        & 0\degree  & 20\degree & 20\degree \\
Perturbation weight     & 0.25      & 0.75      & 0.75      \\
Arm-interarm size ratio & 5         & 5         & 5         \\
Temperature         & 3500 K        & 10000 K       &           \\
Luminosity / mass       & $4\times10^{10}\,\mathrm{L}_\odot$  &  $1\times10^{10}\,\mathrm{L}_\odot$  &  $2\times10^7\,\mathrm{M}_\odot$     \\
    \bottomrule
\end{tabular}
\end{table}

We still assume a homogeneous distribution for the stellar sources and the dust.
The model includes a stellar bulge and disk with an older star population, a second stellar disk with a younger star population, and a dust disk.
The relevant parameters are listed in Table~\ref{tab:SpiralParams}.
The bulge is modeled by a spheroidal density distribution obtained by flattening a spherical S\'ersic profile \citep{Sersic1963}, as implemented by \citet{Baes2015}.
The distributions of the two stellar and the dust disks are truncated double-exponential disks.
The spiral arm structure is introduced by adding a perturbation to the overall density profile, as presented by \citet{Baes2015}.
We have made the spiral arms in the older stellar population `lead' those in the younger stellar population and in the dust disk by varying the spiral arm phase zero-points.
The emission of the stellar populations is modeled as blackbody spectra at the indicated temperatures.
We use the dust model of Sect.~\ref{sec:bianchi} \citep{Camps2015b}.
The total dust mass is given in Table~\ref{tab:SpiralParams}, the central face-on $B$-band optical depth of the dust disk is $\tau_\mathrm{V}\approx1.3$.

Figure~\ref{fig:centralFig} shows surface brightness maps (color scale) overlaid with linear polarization maps (line segments). The leftmost column is for this model at \mum{1}. The top row shows the model face-on, the second row inclined ($53^\circ$), and the third row edge-on. The polarization degree is up to 1\% around the central part of the models and over the whole map the average polarization degree is similar to the average polarization degree from the B/T=0.5 model from before. As in the B/T=0.5 model, the orientation of the polarization is circular around the central bulge, and for the inclined view the polarization degree left and right of the center increases, while it decreases behind and in front of it. In contrast to the azimuthally uniform model, there is a spiral structure in the polarization map. The linear polarization degree is higher in the arm regions and disappears in the interarm region. The maximum polarization degree is slightly inward from the regions of the arms with the highest flux.
The panel in the bottom row plots the linear polarization degree for the edge-on view, averaged over the vertical axis.
This average is obtained by summing each individual component of the Stokes vector and calculating the polarization degree from these totals.
Regions with higher linear polarization (up to 2\%) clearly trace the spiral arms and are prominent at all inclinations, including the edge-on view.
The maxima in the polarization signature of the edge-on view match the positions of the spiral arms along the line of sight.

To verify this, the middle column of Fig.~\ref{fig:centralFig} shows the same model from a different azimuth angle.
The peaks in the polarization signature align with the tangent points of the spiral arms, which are now farther out from the center of the galaxy.

In the rightmost column of Fig.~\ref{fig:centralFig} we remove the spiral arms perturbations from the stellar disks in the model.
The polarization signature remains essentially unchanged; the maxima are slightly higher (by a factor of up to 1.2), but the structure is the same.
The signature could also be produced by the different phase zero-points of the old stars and the dust (see Table~\ref{tab:SpiralParams}). We calculated results using the same phase zero-point for all components (not shown here). The outer maxima are lower (by a factor of 0.8), while the inner maxima are unchanged.
This confirms that the polarization signature is created by the distribution of the dust and not by the distribution of the sources.

In Fig.~\ref{fig:inclinations} we further study the effect of inclination on the observed polarization signature for the reference model.
In the central region ($r\lesssim\kpc{2}$), the bulge emission masks most polarization at inclinations above $40\degree$.
Outside this region, however, the form of the curves is very similar for all inclinations.
Although the polarization degree generally decreases toward lower inclinations, the peaks remain at the spiral arm tangent points, and the ratio between the maximum and minimum polarization degree remains roughly stable at a factor of about 2.

In Fig.~\ref{fig:multWavelength} we compare the edge-on polarization signature of our reference model for various optical and near-infrared wavelengths.
The polarization peaks remain prominent over the wavelength range $\mum{0.5}\lesssim\lambda\lesssim\mum{2}$.
At shorter wavelengths the general polarization degree is higher and the signature is reversed, light from within the arms is slightly less polarized than light from the inter-arm regions. The increased interaction cross section causes the inter-arm dust to become efficient at scattering the stellar radiation, boosting the polarization degree. We find that in the spiral arms the ratio of once scattered to multiple times scattered light is 2.5:1, while in the inter-arm region it is 3.3:1. The polarization orientation after multiple scatterings is less uniform, which lowers the polarization degree in the arm regions.

At longer wavelengths, the signature retains the same form, but the reduced scattering efficiency of the dust causes the polarization degree to be very low, so that the peaks become hard to discern.

Our results imply that polarization measurements could be used, at least in principle, to study the spiral structure of edge-on spiral galaxies, where intensity measurements alone have limited diagnostic power. The contrast would be highest at around \mum{1} and with an expected polarization degree of up to 0.6\%, this is well within the capabilities of current polarization capable telescopes.

We note that the polarization degree of the edge-on galaxy NGC 891 was mapped by \citet{Montgomery2014} at \mum{1.6}. They found polarization degrees of below 1\% that varied along the disk profile. We expect the orientation of polarization due to scattering to be perpendicular to the disk. \citet{Montgomery2014} found the orientation to be rather parallel to the disk and attributed most of it to dichroic extinction. Our code does not yet support dichroic extinction, so we cannot compare the strength of these two effects.


\section{Conclusion}\label{sec:Conclusion}

We presented a robust framework that is independent of a coordinate
system for implementing polarization in a 3D MCRT code, focusing on scattering by spherical dust grains.
The mathematical formulation and the numerical calculations in our method rely solely on the scattering planes determined by the physical processes rather than by the coordinate system.
This approach avoids numerical instabilities for special cases and enables a more streamlined implementation.
We described four test cases with well-defined geometries and material properties, yielding analytical solutions.
These setups are designed to validate the calculated results in a structured manner, and they can serve as benchmarks for other implementations as well.

We reconstructed a selection of the 2D models of \citet{Bianchi1996}, confirming that our implementation reproduces their results at least qualitatively. A quantitative comparison is not possible because of differences in the dust model.
We then calculated results for the polarization part of the \citet{Pinte2009} benchmark and obtained similar results as the other codes that took part in it.
As an application of our code we constructed a 3D spiral galaxy model including a stellar bulge and disk with an older star population, a second stellar disk with a younger star population, and a dust disk.
The stellar and dust distributions feature an analytical spiral arm perturbation.
We showed that scattering of light at the dust in the spiral arms produces a marked polarimetric signature. It traces the tangent positions of the arms for wavelengths in the range $\mum{0.5}\lesssim\lambda\lesssim\mum{2}$, regardless of inclination.

It is fair to note, however, that our current implementation is limited to scattering by spherical dust grains. We plan to add support for polarized emission and for scattering and absorption by (partially) aligned spheroidal grains in future work. With the relevant physics included, we can also study the influence of changes to dust properties on the polarization signature, which we have not addressed in this paper.

\begin{acknowledgements}
C.P., P.C., and M.B. acknowledge the financial support from CHARM (Contemporary physical challenges in Heliospheric and AstRophysical Models), a Phase-VII Interuniversity Attraction Pole program organized by BELSPO, the BELgian federal Science Policy Office. M.S. acknowledges support by FONDECYT through grant no. 3140518 and by the Ministry of Education, Science and Technological Development of the Republic of Serbia through the projects Astrophysical Spectroscopy of Extragalactic Objects (176001) and Gravitation and the Large Scale Structure of the Universe (176003).
We thank Rene Goosmann, Francesco Tamborra, and Frederic Marin for many useful discussions and providing insights on the operation of the STOKES code. We wish to thank the anonymous referee for their constructive input to this paper.
\end{acknowledgements}
\bibliographystyle{aa} 
\bibliography{Library}


\end{document}